\documentclass{aa}  

\usepackage{natbib}

\usepackage{graphicx}
\usepackage{subcaption}
\usepackage{multirow} 
\usepackage{xcolor}
\usepackage[normalem]{ulem}
\usepackage{booktabs}
\usepackage{txfonts}
\newcommand{\scriptfootnotesize}{\fontsize{8.2pt}{10pt}\selectfont}
\usepackage[colorlinks=true,linkcolor=blue,citecolor=blue,urlcolor=blue]{hyperref}
\begin{document}

   \title{Spatial variability of CH$_4$ and C$_2$H$_2$ absorptions in Jupiter’s auroral regions from Juno-UVS observations}

   \author{J. Vinesse\inst{1} \and
          B. Bonfond\inst{1} \and
          B. Benmahi\inst{2,1} \and
          A. Moirano\inst{1,3} \and
          D. Grodent\inst{1} \and
          T.K. Greathouse\inst{4} \and
          V. Hue\inst{2} \and
          G. Sicorello\inst{1} \and
          L.A. Head\inst{1} \and
          G.R. Gladstone\inst{4} \and
          M.W. Davis\inst{4}
          }

   \institute{Laboratory for Planetary and Atmospheric Physics, STAR Institute, University of Liège, Liège, Belgium\\
              \email{julie.vinesse@uliege.be}
        \and
             Aix-Marseille Université, CNRS, CNES, Institut Origines, LAM, Marseille, France
        \and   
             Institute for Space Astrophysics and Planetology, National Institute for Astrophysics (INAF-IAPS), Rome, Italy
        \and
             Southwest Research Institute, San Antonio, Texas, USA
             }

   \date{}

% \abstract{}{}{}{}{} 
% 5 {} token are mandatory
 
  \abstract
   {Ultraviolet auroral emissions on Jupiter offer a unique window into the coupling between the planet's magnetosphere and upper atmosphere. Color ratios derived from molecular hydrogen emissions provide valuable diagnostics for the energy of precipitating electrons and the structure of the auroral atmosphere.}
   {We aim to characterize the horizontal and vertical variability of hydrocarbon absorption in Jupiter’s auroral atmosphere using ultraviolet data from the Juno-UVS spectrograph and to investigate potential departures from the expected structure.}
   {We constructed color ratio maps sensitive to CH$_4$ and C$_2$H$_2$ absorptions for perijoves (PJs) 6 and 10, two of Juno’s close approaches to Jupiter, by integrating auroral H$_2$ emission over hydrocarbon-sensitive spectral intervals. For CH$_4$, we redefined the absorbed spectral band, replacing the traditionally used 125–130 nm interval with 135–140 nm, in order to mitigate higher-order calibration issues. In regions of intense auroral brightness, we developed a correction method to account for spectral distortion due to detector non-linearities at high fluxes. }
   {CH$_4$ and C$_2$H$_2$ absorptions generally follow the expected vertical distribution, with the CH$_4$ density extending to higher altitudes than C$_2$H$_2$. However, several localized regions show unexpected spatial distribution of the absorption. In PJ6, such anomalies are attributed to instrumental non-linearities. After correction, the CR distributions become consistent with standard hydrocarbon vertical distributions. In PJ10, however, some anomalous patterns persist despite correction. Spectral modeling indicates that these can be explained by modifying the relative abundances of CH$_4$ and C$_2$H$_2$, suggesting horizontal compositional variability and possible deviations in homopause altitude between species.}
   {Our results confirm that hydrocarbon absorption in Jupiter’s aurora is vertically stratified, but that the altitude of this stratification itself varies horizontally with latitude and longitude. These findings underscore the importance of accounting for local atmospheric composition when interpreting ultraviolet auroral spectra and retrieving electron energy distributions. Juno-UVS thus provides a powerful diagnostic for probing both auroral precipitation and upper atmospheric structure. The framework developed here enables more accurate retrievals of electron energies and provides new constraints on the spatial variability of Jupiter's upper atmosphere. A systematic application to the full Juno dataset will offer deeper insights into the temporal and spatial dynamics of the Jovian aurora.}

   \maketitle
%
%-------------------------------------------------------------------

\section{Introduction}

Jupiter’s auroral emissions mostly originate from the precipitation of energetic electrons that have been accelerated within the planet’s magnetosphere. Upon entering the upper atmosphere, these electrons collide with atmospheric constituents, leading to excitation, dissociation and ionization processes that generate characteristic ultraviolet emissions. As such, auroral emissions offer a remote diagnostic of magnetospheric dynamics and provide key insights into the physical mechanisms responsible for electron acceleration.

The first ultraviolet detection of Jovian aurorae was achieved during the Voyager 1 flyby, which revealed emissions from both atomic hydrogen (Lyman-$\alpha$) and molecular hydrogen (H$_2$) bands, specifically the Lyman (B$^1\Sigma^+_u \rightarrow$ X$^1\Sigma^+_g$) and Werner (C$^1\Pi_u \rightarrow$ X$^1\Sigma^+_g$) transitions \citep{broadfoot_extreme_1979}. This discovery marked the beginning of several decades of continuous monitoring, notably by the International Ultraviolet Explorer (IUE), which provided valuable temporal coverage and spectral information \citep{Clarke1980, Livengood1992, Gladstone1989, Harris1996}.

More recently, ultraviolet observations conducted with the Hubble Space Telescope (HST) have considerably enhanced our understanding of the aurora. Imaging instruments such as the Faint Object Camera (FOC) and the Wide Field and Planetary Camera 2 (WFPC2) have offered high-resolution views of auroral morphology \citep{Dols1992, Gerard1994, Prange1998, Clarke1996, Clarke1998, Grodent1997}, while the Space Telescope Imaging Spectrograph (STIS) has provided complementary spectroscopic diagnostics \citep{Gustin2002}. Despite the wealth of data collected, the spatial and temporal complexity of Jupiter’s auroral structures continues to challenge a complete characterization \citep{grodent_brief_2015}. Nonetheless, these observations have enabled a general classification of the main auroral components: a bright, partially closed main emission; dynamic emissions located poleward of this structure; and equatorward features including the auroral footprints of the Galilean moons Io, Europa and Ganymede.

Beyond spatial morphology, spectroscopic measurements provide a powerful means of probing the energy of the precipitating electrons responsible for auroral emissions. In particular, the color ratio (CR) has proven to be a robust diagnostic for constraining the depth of electron penetration into Jupiter’s upper atmosphere.

Originally introduced by \citet{yung_h2_1982}, the CR exploits the wavelength-dependent absorption of H$_2$ emission by atmospheric hydrocarbons, primarily methane (CH$_4$). It is defined as the ratio of auroral brightness in a weakly absorbed spectral interval (155–162~nm) over that measured in a strongly absorbed interval.
 In its earliest formulation, the denominator spanned the 123--130 nm range, but this was later refined to 125--130 nm to avoid contamination by the instrumentally broadened Lyman-$\alpha$ line when using the UltraViolet Spectrograph (UVS) \citep{gerard_contemporaneous_2019}. The currently adopted form is thus

\begin{equation*}
    CR = \frac{I(155\text{--}162~\text{nm})}{I(125\text{--}130~\text{nm})}.
\end{equation*}

This ratio provides a physically meaningful diagnostic because CH$_4$ is confined below the homopause, the transition altitude between molecular diffusion and eddy diffusion. Below this level, heavier hydrocarbons are well-mixed due to turbulent mixing; above it, molecular diffusion dominates and their densities drop off sharply with altitude. Auroral H$_2$ emissions generated above the homopause (by low-energy electrons) undergo little absorption, whereas emissions produced deeper in the atmosphere (by higher-energy electrons) are increasingly attenuated by CH$_4$. Consequently, a higher CR indicates deeper penetration of energetic particles and thus a higher characteristic electron energy.

Alternative formulations of the CR can be used to probe the influence of other hydrocarbons. For example, for instruments with limited spectral coverage at longer wavelengths, such as Hisaki/EXCEED (which extends only up to 148~nm), a modified ratio can be adopted. \citet{tao_variation_2016} define CR$_\text{EXCEED}$ as
\begin{equation*}
    \mathrm{CR}_{\text{EXCEED}} = \frac{I(138.5\text{--}144.8~\text{nm})}{I(126.3\text{--}130~\text{nm})} \times 1.10,
\end{equation*}
where the chosen bands ensure differential CH$_4$ absorption, avoid H$_2$ self-absorption and fall within the instrument’s sensitivity range. This example illustrates how the CR framework can be adapted to different spectral instruments while retaining its diagnostic value.

A quantitative link between the CR and the characteristic energy of precipitating electrons has long been exploited as a remote sensing tool to constrain electron penetration depths in planetary aurorae \citep[e.g.,][]{trafton_fuv_1994, Grodent2001, Gustin2002, ajello_properties_2005, gerard_mapping_2014, gustin_characterization_2016}. This technique relies on the fact that CH$_4$ resides predominantly below the homopause, so that increasing absorption in CH$_4$-sensitive spectral bands reflects deeper energy deposition and thus higher-energy electrons. The CR offers a relationship with the characteristic energy $E_0$ of precipitating electrons, provided the emission geometry is fixed.

Building on this established principle, \citet{benmahi_energy_2024} presented a refined version of the CR–$E_0$ relationship using the \texttt{TransPlanet} \citep{lilensten_ionization_1989} electron transport model. This model combines electron transport and H$_2$ excitation with detailed UV radiative transfer through a hydrocarbon-rich atmosphere and includes the effect of photon emission angle, which modulates the atmospheric path length and thus the probability of absorption along the line of sight. By simulating auroral emissions for both monoenergetic and kappa electron energy distributions and computing the corresponding synthetic CRs, a calibration curve was derived, which can be inverted to estimate $E_0$ from observed CRs.

Compared to in-situ particle measurements from instruments such as JADE and JEDI onboard Juno, this spectroscopic method offers two main advantages. First, UVS spectro-imaging provides much broader spatial and temporal coverage than point-sampling particle detectors, enabling a global view of auroral precipitation. Second, in-situ instruments sometimes probe electrons above the acceleration region (~1-2 R$_J$, where R$_J$ is Jupiter’s equatorial radius, 71,492 km) and may not reflect their energy at atmospheric impact. The spectroscopic approach, by contrast, diagnoses the actual energy deposition in the upper atmosphere and therefore more directly constrains the energy of the precipitating electrons but relies on models and assumptions.

This spectroscopic method, though powerful, inherently depends on the assumed vertical and horizontal distribution of CH$_4$ in the atmosphere. In particular, any variation in the altitude of the homopause, above which CH$_4$ is strongly depleted, can significantly impact the absorption depth of UV photons and therefore bias the energy retrieval from CR measurements \citep{benmahi_auroral_2024}.

Infrared observations, particularly from Juno and ground-based telescopes, have revealed vertical variability up to 100 km in the altitude of CH$_4$ and other hydrocarbons across Jupiter’s auroral regions \citep{sinclair_2017, sinclair_2018, sinclair_2020, sinclair_2025, giles_enhanced_2023}. These compositional gradients likely result from magnetospheric particle precipitation, which alters both the chemical composition \citep{sinclair_2019, sinclair_2023, benmahi_monitoring_2020, cavalie_evidence_2023} and thermal structure \citep{o2021global} of the upper atmosphere (see \citet{hue2024polar} for a comprehensive review).

In particular, variations in the altitude of the CH$_4$ homopause have been observed and can reach 100 km between the regions poleward and equatorward of the main emission \citep{sinclair_2025}. Since the CR depends critically on the amount of CH$_4$ encountered by auroral photons-i.e., on the depth at which UV emissions are absorbed-such variability directly affects the inferred energy of precipitating electrons. For high-energy electrons, which deposit their energy deeper in the atmosphere, a higher homopause leads to increased absorption and thus a higher CR, potentially resulting in an overestimate of the characteristic energy $E_0$ if not properly accounted for. Conversely, a lower homopause altitude leads to less CH$_4$ absorption and may cause an underestimation of $E_0$.

This effect can introduce systematic uncertainties of up to 200~keV in the inferred energies of the most energetic auroral electrons, as shown in \citet{benmahi_auroral_2024}. Therefore, any attempt to retrieve accurate particle energy distributions from CR measurements must consider the horizontal variability of CH$_4$ and the vertical structure of the atmosphere.
In this study, we refine hydrocarbon absorption diagnostics in Juno-UVS data, apply them to PJ6 and PJ10, and reveal new insights into the spatial distribution of CH$_4$ and C$_2$H$_2$ in Jupiter’s auroral atmosphere.

\section{Instrumentation and models}
\subsection{The Juno Mission and the UVS Instrument}

The Juno spacecraft, part of NASA’s New Frontiers program, entered orbit around Jupiter on 4 July 2016 following its launch in 2011. The mission's primary objectives include probing Jupiter’s internal structure, magnetic and gravitational fields, atmospheric composition and dynamics, and investigating the nature and drivers of its auroral emissions \citep{Bolton2017, Bagenal2017}.

To fulfill these goals, Juno was placed on a highly elliptical polar orbit designed to minimize radiation exposure while enabling close approaches to the planet. Each orbit, referred to as a perijove (PJ), brings the spacecraft within approximately 4200~km of the cloud tops and extends out to over 100~R$_\mathrm{J}$ at apoapsis. 

Among Juno’s nine scientific instruments is UVS, a key tool for auroral studies. UVS is a far- and extreme-ultraviolet spectrograph covering the 68–210~nm range \citep{Gladstone2017}. It uses a flat scanning mirror to direct incoming photons toward a primary mirror and a dogbone-shaped slit, composed of two wide segments at the ends (0.2$^\circ \times$ 2.5$^\circ$) and a narrower central segment (0.025$^\circ \times$ 2.0$^\circ$). Light passing through the slit is dispersed by a diffraction grating and detected on a two-dimensional sensor.

The spin-stabilized spacecraft completes one rotation every $\sim$30 seconds, allowing UVS to sweep across the auroral regions and build up spatial coverage. Each rotation yields spectrally resolved photon counts associated with wavelength, time and emission angle. 

The UV emissions observed by UVS primarily originate from atomic hydrogen (Lyman-$\alpha$) and H$_2$ transitions in the Lyman and Werner bands. These features result from electron-impact excitation of the upper atmosphere and provide critical diagnostics of auroral processes. For this study, we use UVS observations to investigate spectral and energetic characteristics of Jupiter’s aurora.
\subsection{Atmospheric Model: Grodent et al. (2001)}

The structure of the auroral atmosphere assumed in this study is based on the 1D model described by \citet{Grodent2001}. This model provides vertical profiles for the major neutral species in Jupiter’s upper atmosphere (H, H$_2$, He, CH$_4$ and C$_2$H$_2$) from the tropopause (near 100~mbar, approximately 50~km above the visible cloud tops) to the upper thermosphere ($10^{-9}$~mbar, or about 2300~km altitude).

Hydrogen (both molecular and atomic) and helium are present at all altitudes, while CH$_4$ and C$_2$H$_2$ are mostly confined below the homopause. 

The adopted model assumes a horizontally homogeneous atmosphere, which is a simplification. As discussed in the introduction, infrared observations have revealed significant variability in the altitude of the CH$_4$ homopause across auroral latitudes, which can impact the reliability of CR–based energy diagnostics.

Despite this limitation, the \citet{Grodent2001} model remains widely used in auroral studies due to its physical consistency and compatibility with radiative transfer codes such as \texttt{TransPlanet} and it provides a reasonable baseline for investigating first-order effects in the auroral regions.

\subsection{Electron transport and emission: TransPlanet}

To simulate electron precipitation and the resulting ultraviolet emissions in Jupiter’s auroral atmosphere, we use the TransPlanet electron transport model. Originally developed by \citet{lilensten_ionization_1989}, this model has since been adapted to various planetary environments, including studies of Jupiter’s auroral regions and the interaction between Triton’s atmosphere and Neptune’s magnetosphere \citep{benmahi_etude_2022,benmahi_energy_2024,benmahi_auroral_2024,benne_impact_2024}. TransPlanet also includes a dedicated UV emission model to compute H$_2$ emissions in the 80~nm, 210~nm range generated by excitation from precipitated magnetospheric electrons collisions into Jupiter’s atmosphere \citep{benmahi_energy_2024, benmahi_auroral_2024}. 

TransPlanet solves the Boltzmann equation for each altitude bin for a suprathermal electron population precipitating into a model atmosphere. The equation accounts for both elastic and inelastic collisions with atmospheric species, including the production of secondary electrons. The Jovian atmospheric profile used is based on \citet{Grodent2001}.

After solving the transport problem, the resulting altitude-dependent electron energy flux is used as input for the UV emission model, which calculates the excitation and subsequent radiative de-excitation of H$_2$ molecules. This process produces ultraviolet emissions primarily in the Lyman and Werner bands. The UV emission model includes direct excitation, cascade contributions and self-absorption effects, with photon redistribution toward lower-energy transitions \citep{benmahi_energy_2024}. Absorption by hydrocarbons such as CH$_4$, C$_2$H$_2$ and C$_2$H$_6$ is also incorporated to yield realistic spectra. This UV emission model has been validated by \citet{benmahi_energy_2024} against laboratory data, notably the electron-impact excitation spectra of H$_2$ from \citet{Liu1995}, showing good agreement in line positions and relative intensities. 

The final output from TransPlanet includes the total vertical energy flux profile of precipitating electrons and a synthetic UV spectrum by H$_2$. This model spectrum serves as a basis for comparison with UVS observations in Section~\ref{casestudies}.

\section{Data processing and Method}
\subsection{Spectral Cubes}
We constructed spectral cubes for PJ6 and PJ10 using the same method as described in \citet{benmahi_energy_2024}. The input data consist of time-tagged photon lists recorded by UVS, which scans the sky as the spacecraft spins, enabling coverage of Jupiter’s auroral regions in the far-ultraviolet (FUV) range \citep{bonfond2017morphology}. Each detected photon is associated with ancillary parameters such as its time of detection, wavelength (from detector position), detector coordinates (x, y), emission angle and the corresponding projected location on Jupiter in System~III coordinates (latitude and longitude), derived from the spacecraft’s orientation and the instrument’s pointing geometry \citep{ACTON20189, ACTON199665}.

Photon counts are converted into physical units of spectral flux using the instrument's effective area, which is determined from regular stellar calibration observations performed throughout the mission \citep{hue2021updated, hue_-flight_2018, greathouse_performance}. To ensure consistent spectral resolution and maximize the S/N, only photons collected through the two wide slits of UVS are retained for analysis. These wide slits offer a spectral resolution of approximately 2.1~nm \citep{greathouse_performance}. 

The calibrated photon lists are then binned into three-dimensional spectral cubes defined over latitude, longitude (both in System~III) and wavelength. The spatial grid is sampled every $1^\circ$ in latitude and longitude. In the spectral dimension, we adopt a sampling of 0.1~nm, which satisfies the Nyquist criterion based on the UVS spectral point spread function. Photons are accumulated over selected time intervals. The resulting cubes span the 125–170~nm range and provide spatially and spectrally resolved maps of the auroral emission for each hemisphere.

From the spectral cubes, we derive brightness maps of Jupiter's northern and southern auroral regions. In particular, we compute H$_2$ emission maps by integrating the flux over the 155–162~nm wavelength band, which is minimally affected by hydrocarbon absorption and corresponds to the strongest Lyman and Werner band emissions. These maps reveal the large-scale morphology of the auroral emissions, including specific features such as the main emission, the polar emissions and transient features like dawn storms or injection signatures.

\begin{figure}[ht]
    \centering
    % Sous-figure (a) PJ6
    \begin{subfigure}{\linewidth}
        \centering
        \includegraphics[width=\linewidth]{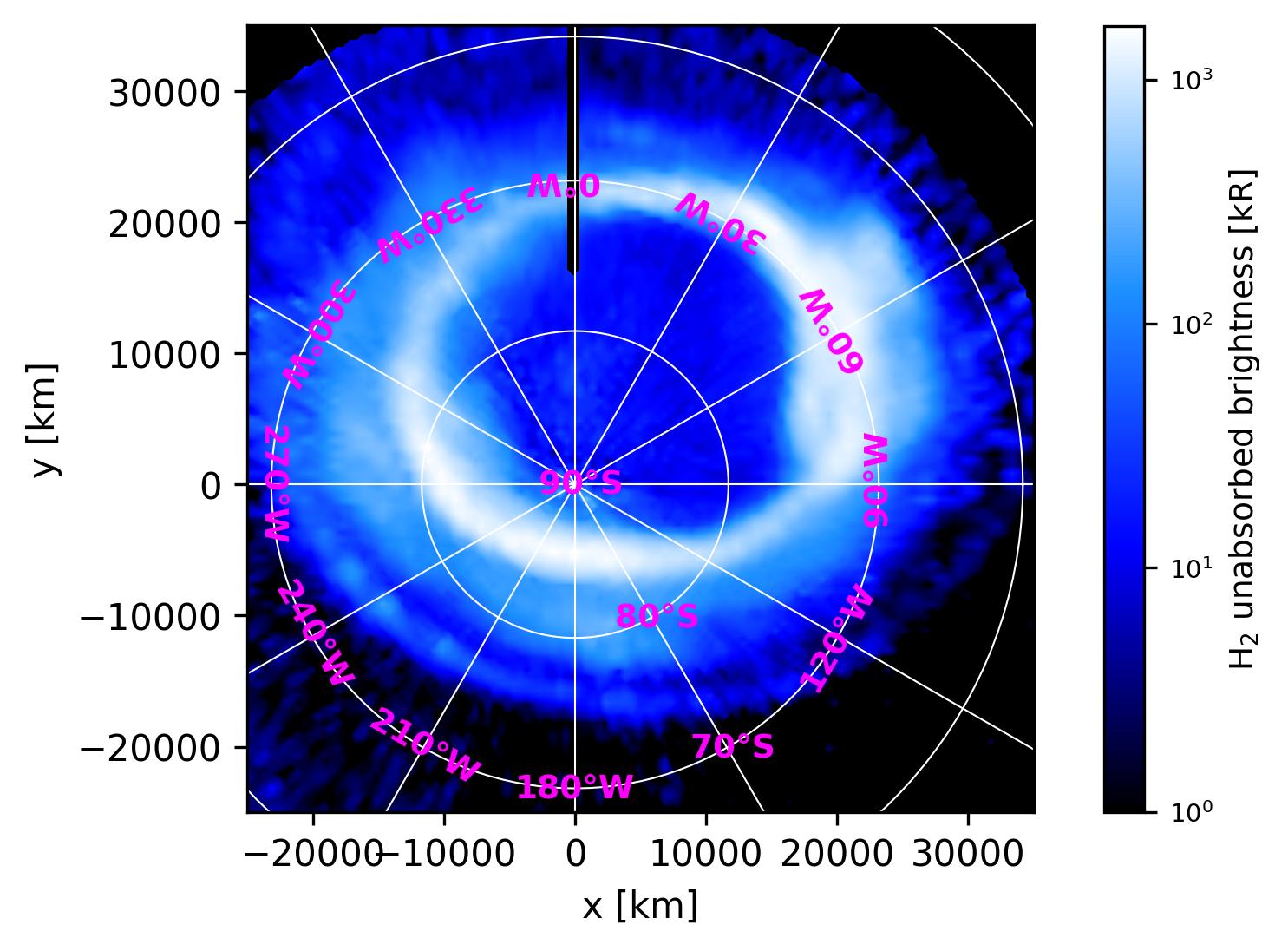}
        \caption{PJ6: H$_2$ brightness map integrated over the 155--162~nm unabsorbed band for the southern aurora.}
        \label{fig:h2_brightness_map_6}
    \end{subfigure}
    
    \vspace{0.4cm}
    
    % Sous-figure (b) PJ10
    \begin{subfigure}{\linewidth}
        \centering
        \includegraphics[width=\linewidth]{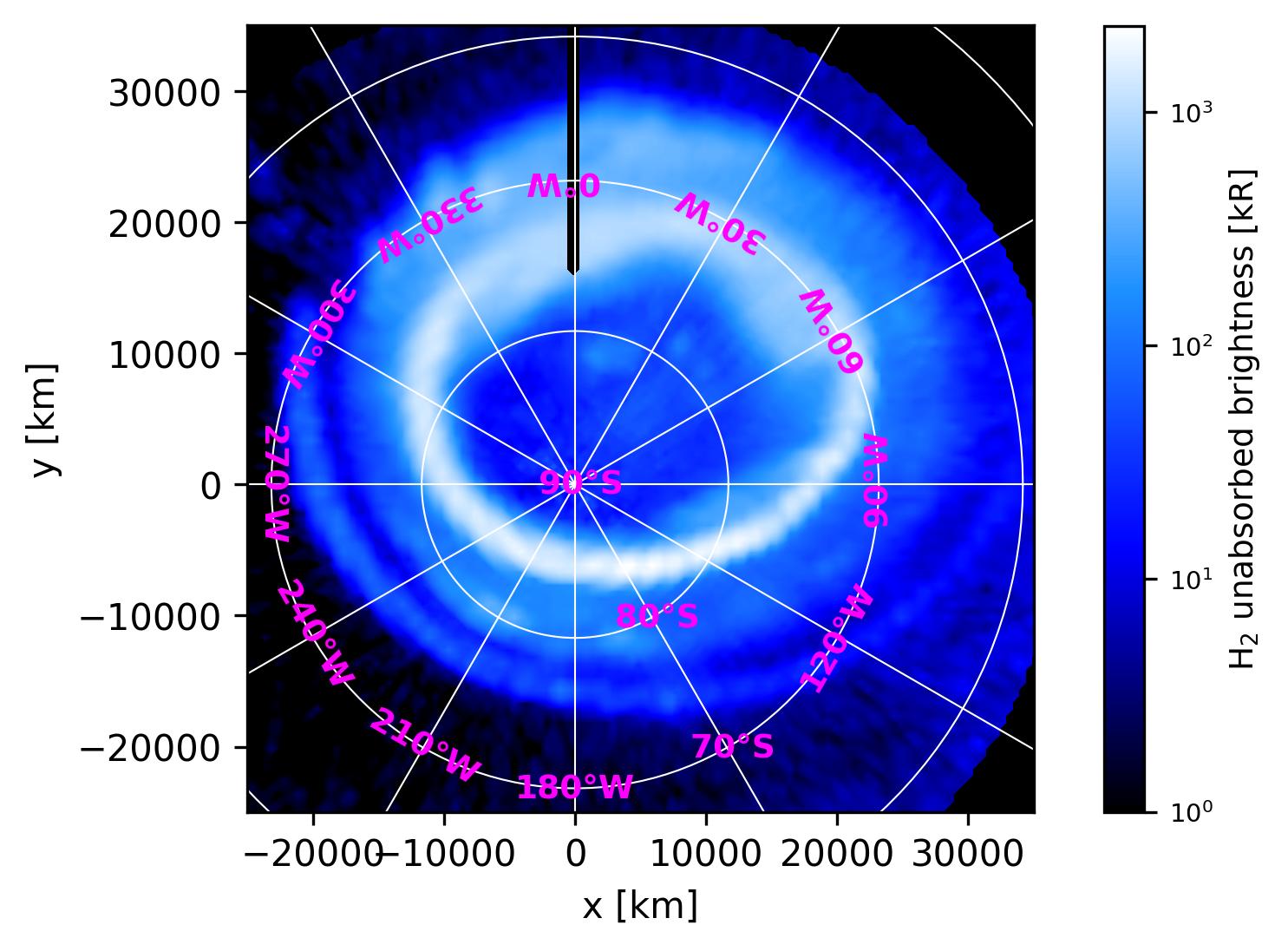}
        \caption{PJ10: Same as (a) for a different perijove.}
        \label{fig:h2_brightness_map_10}
    \end{subfigure}
    
    \caption{Southern auroral H$_2$ brightness maps integrated over the 155--162~nm unabsorbed band for two perijoves: 
    (a) PJ6 and (b) PJ10.}
    \label{fig:h2_brightness_maps}
\end{figure}

Figs.~\ref{fig:h2_brightness_map_6} and \ref{fig:h2_brightness_map_10} show the H$_2$ brightness maps obtained during PJ6 (from 2017-May-19 01:01:05 UTC to 2017-May-19 10:39:48 UTC) and PJ10 (from 2017-Dec-16 13:45:18 UTC to 2017-Dec-16 22:57:00 UTC) for the southern pole. The main auroral emission appears as a bright arc encircling the pole, while all the emissions contained inside of the main emission are called the polar emissions. The emissions outside of the main emission contain signatures of injections as well as the footprints of the jovian satellites. 

The spectral cubes also enable the construction of hydrocarbon absorption maps based on CRs. In particular, we aim to compare the absorptions by CH$_4$ and by C$_2$H$_2$, two hydrocarbons that dominate the optical depth in distinct ultraviolet intervals. Each CR is defined as the ratio of the integrated H$_2$ brightness in an unabsorbed spectral band to that in a hydrocarbon-sensitive absorption band.

For CH$_4$, the CR is defined as
\begin{equation*}
\mathrm{CR}_{\mathrm{CH}_4} = \frac{I(155\text{--}162~\mathrm{nm})}{I(125\text{--}130~\mathrm{nm})},
\end{equation*}
where the 125--130~nm range corresponds to strong CH$_4$ absorption and the 155--162~nm band is only weakly affected by hydrocarbons.

To probe C$_2$H$_2$ absorption, we define
\begin{equation*}
\mathrm{CR}_{\mathrm{C}_2\mathrm{H}_2} = \frac{I(155\text{--}162~\mathrm{nm})}{I(150\text{--}153~\mathrm{nm})},
\end{equation*}
where the 150--153~nm interval is particularly sensitive to C$_2$H$_2$.

\begin{figure}[ht]
    \centering

    % ===== Spectre TransPlanet =====
    \hspace*{0.2mm}
    \begin{subfigure}[b]{\columnwidth}
        \centering
        \includegraphics[width=\linewidth]{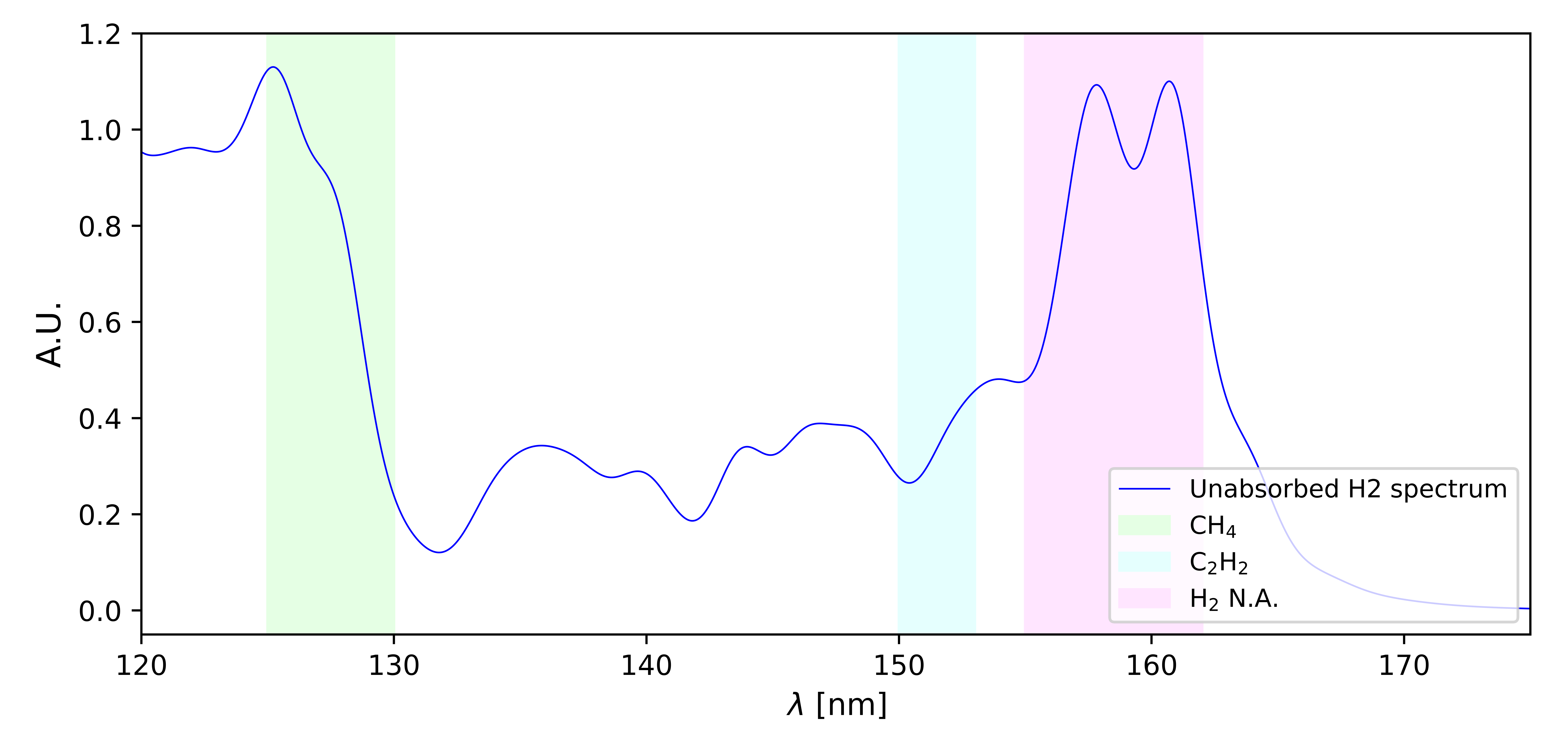}
        
        \label{fig:transplanet_h2_spectrum}
    \end{subfigure}

    \vspace{0.2cm}

    % ===== Optical depth Benmahi 2024 =====
    \begin{subfigure}[b]{\columnwidth}
        \centering
        \includegraphics[width=1.036\linewidth]{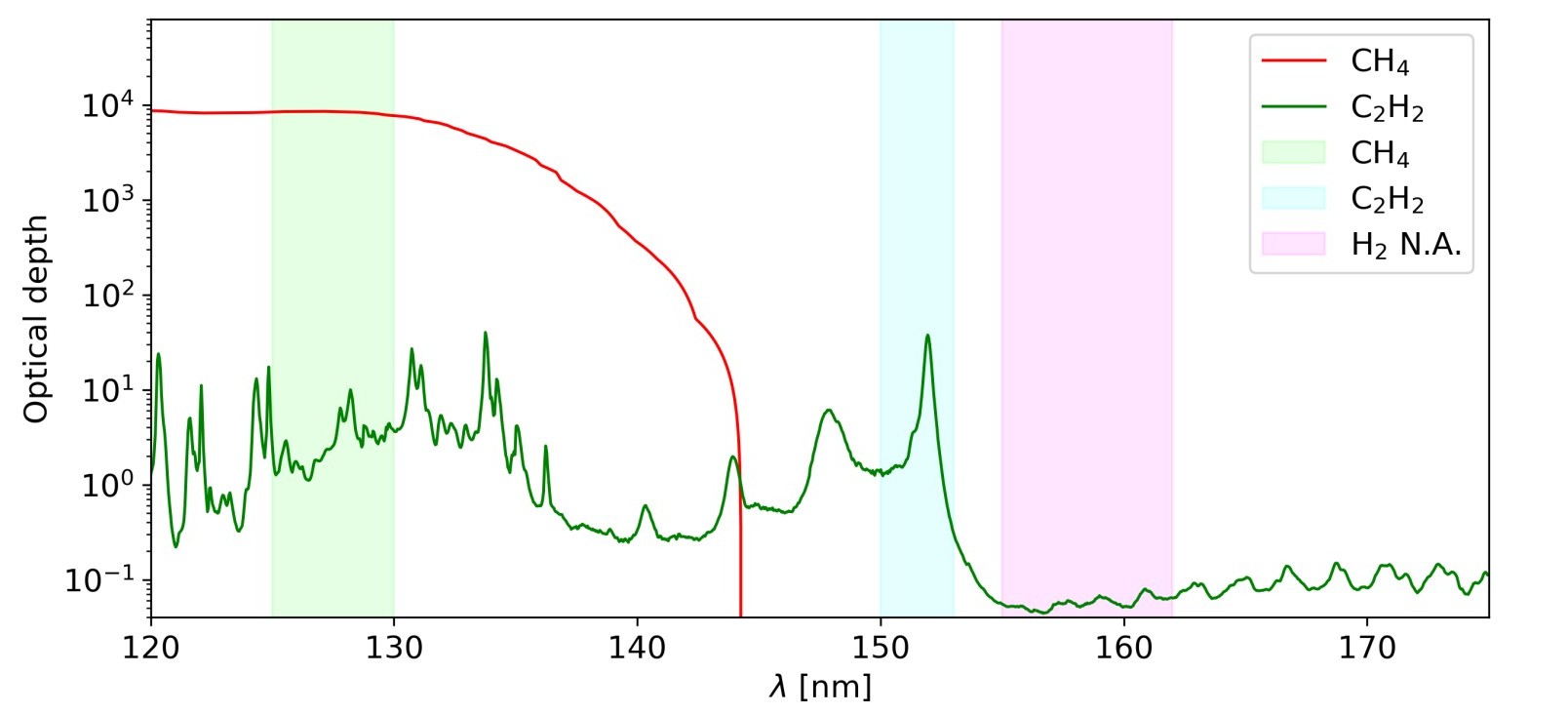}
        \label{fig:optical_depth_cr_bands_benmahi}
    \end{subfigure}

    \caption{Spectral diagnostics for hydrocarbon absorption and CR computation. Transparent green and cyan bands indicate the absorption spectral ranges used for CR calculations (125--130\,nm for CH$_4$, 150--153\,nm for C$_2$H$_2$), while the magenta band represents the non-absorbed region assumed to be free of hydrocarbon absorption. Upper panel: Simulated unabsorbed H$_2$ spectrum from TransPlanet in dark blue. Lower panel: Corresponding optical depth for CH$_4$ and C$_2$H$_2$, showing the ranges used in CR calculations (from \citet{benmahi_energy_2024}).}
    \label{fig:spectrum_and_optical_depth_column}
\end{figure}

These ratios allow us to isolate the contribution of each hydrocarbon species. Fig.~\ref{fig:spectrum_and_optical_depth_column} illustrates the spectral intervals used for the CR calculations: panel (a) shows a simulated unabsorbed H$_2$ spectrum from TransPlanet, with absorption bands indicated, while panel (b), from \citet{benmahi_energy_2024}, displays the optical depth for CH$_4$ and C$_2$H$_2$ highlighting the same spectral ranges.
In the absence of strong dynamical or chemical perturbations, CR$_{\rm CH_4}$ and CR$_{\rm C_2H_2}$ maps are expected to exhibit similar morphologies. This expectation is justified because both CH$_4$ and C$_2$H$_2$ are primarily located below their respective homopauses, where molecular diffusion is slow and advective transport dominates, leading to comparable spatial distributions.

\subsection{Detector Response at High Flux Levels}
The photon detection process is characterized by the pulse height distribution (PHD), which characterizes the distribution of output signal amplitudes generated by photoelectron avalanches in the microchannel plates (MCPs) of the UVS detector \citep{hue_-flight_2018, greathouse_performance}. When a UV photon strikes the photocathode, it ejects a photoelectron that initiates a cascade of secondary electrons within the MCP channels. Each such cascade results in an electronic pulse and the amplitude (height) of these pulses depends on several factors, including gain uniformity, electron multiplication statistics and the intrinsic variability of the MCP.

A PHD that displays as a unimodal distribution centered at above a typical amplitude is indicative of stable gain and reliable photon counting. However, when observing very bright regions or when using instrumental configurations that increase the incoming flux (e.g. the wide slit), the PHD can become distorted. This distortion results in a shift of the peak to lower pulse heights and a broadening or flattening of the distribution. In practice, such degraded PHDs result in undercounting of detected photons.

\begin{figure}[h]
\centering
\includegraphics[width=\linewidth]{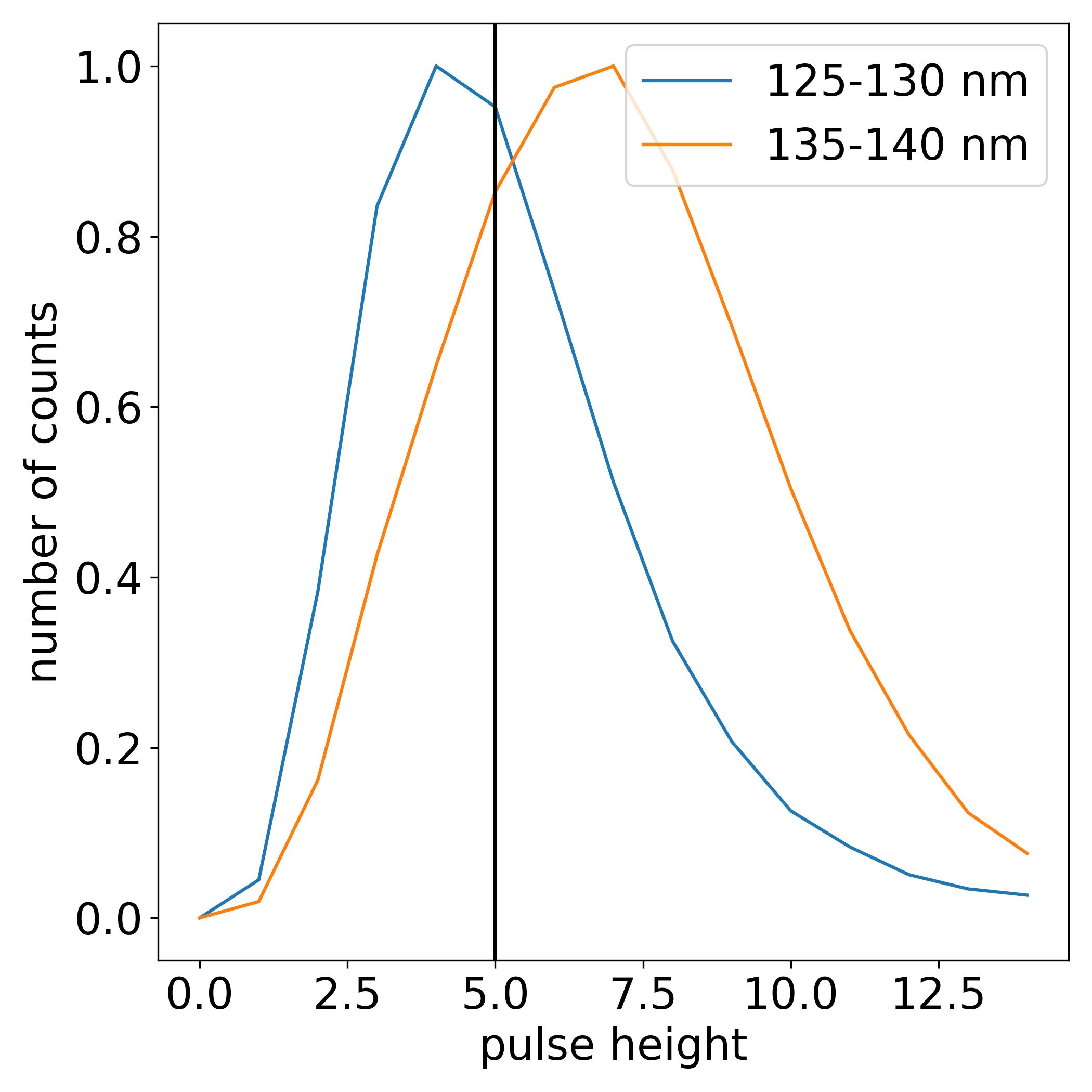}
\caption{Comparison of PHDs in the 125–130~nm spectral band (blue) and the 135-140 nm spectral band (orange) during PJ10 for the region between 70°S and 73°S and between 30°W and 60°W. The orange histogram (135-140 nm) shows a well-calibrated distribution with a peak above 5. The blue histogram (125-130 nm) shows a PHD with a degraded shape and a peak below 5, indicating significant signal loss.}
\label{fig:phd_comparison}
\end{figure}

This effect is illustrated in Fig.~\ref{fig:phd_comparison}, which compares two PHDs recorded during the PJ10 flyby over the same auroral region: one in the 135-140 nm interval, yielding a normal PHD with a peak above 5 (which is the acceptable threshold) and one in the 125-130 m band, where the PHD is significantly degraded with the peak falling below the acceptable threshold. 

This instrumental effect is spectrally dependent. Before constructing the CR maps, we therefore assess the reliability of the pulse height distribution (PHD) in different spectral intervals. The detector regions receiving the highest photon flux—such as those associated with the Lyman-$\alpha$ line and the unabsorbed band (155–162~nm)—are the most susceptible to high flux-induced distortion of the PHD.

The commonly used CH$_4$ absorption band at 125–130~nm lies close to the intense Lyman-$\alpha$ emission, which is known to suffer from instrumental scrubbing causing gain sag. 
To assess the impact of this effect, we performed a Pulse Height Distribution (PHD) analysis for the 125--130~nm range, the results of which are presented in Appendix~\ref{app:crband}. This analysis shows that, in many regions, the PHD peak falls below the critical threshold of 5, indicating that the detector response deviates from the nominal calibration and that the corresponding measurements are likely unreliable.

\subsection{Revising the CR Definition}
\label{sec:cr_definition}

To mitigate the impact of these calibration issues, we redefine the absorbed band used in the CH$_4$ CR calculation. We select the 135–140~nm interval as a more robust alternative (see Fig.\ref{fig:phd_comparison}), which remains within the CH$_4$ absorption range but is less affected by instrumental artifacts. A more detailed insight into the choice of this new interval is provided in Appendix \ref{app:crband}.

Accordingly, we adopt the revised definition
\begin{equation}
\text{CR}_{\text{CH}_4} = \frac{I(155\text{--}162~\text{nm})}{I(135\text{--}140~\text{nm})}
\label{eq:new_cr_txt}
\end{equation}
This updated definition improves the consistency of our analysis and ensures more accurate diagnostics of hydrocarbon absorption in the auroral regions.

\subsection{Modeling the CR–Energy Relationship with TransPlanet}
\label{sec:transplanet}
To interpret the revised CH$_4$ CR as a diagnostic of precipitating electron energy, we derive an empirical relationship between the CR and the characteristic energy $E_0$ of the electron distribution. Our approach follows the methodology developed by \citet{benmahi_energy_2024}, who established a two-dimensional functional dependence $\mathrm{CR}(E_0, \theta)$ using the \texttt{TransPlanet} code.

In our case, we apply this same method to the newly adopted CH$_4$ CR (Eq. \ref{eq:new_cr_txt}). We simulate H$_2$ auroral emission spectra using \texttt{TransPlanet} for both monoenergetic and kappa-type electron energy distributions, over a grid of $E_0$ values and emission angles $\theta$ (from 0° to 80°). Each spectrum is convolved with the Juno-UVS instrument line-spread function and the CR is computed accordingly. 

Following the formalism proposed by \citet{benmahi_energy_2024}, the CR–energy–angle relationship is fitted using the following analytical expression

\begin{align}
\text{CR}(E_0, \theta) =\ 
& \left[
A \cdot \left(\tanh\left(\frac{E_0 - E_c}{B \cdot (1 - C \cdot \sin(\theta))}\right) + 1\right)
\right. 
\label{eq:cr_model} \\[-0.2em]
& \left. \cdot \log\left(\left(\frac{E_0}{D}\right)^{\alpha} + e\right)^{\beta}
\right] \notag
\\[-0.2em]
& \cdot \left[
1 + \delta \cdot \left(\sin(\theta)\right)^{\gamma / (E_0 / 1.5 \times 10^5)}
\right] \notag
\end{align}

where $E_0$ is the characteristic electron energy (in eV), $\theta$ is the emission angle, $A$ is the amplitude scaling factor, $E_c$ is a threshold energy (in eV), $B$ and $C$ modulate the smoothness and angular sensitivity of the tanh transition, $D$, $\alpha$ and $\beta$ define the logarithmic energy growth at high energies, $\delta$ and $\gamma$ account for angular modulation at high energies, $e$ is Euler's number, ensuring continuity at low energy. The fitting procedure uses a Markov Chain Monte Carlo (MCMC) optimization with the \texttt{emcee} sampler \citep{foreman2013emcee}, using 250 walkers and 2500 iterations to ensure convergence. 

The best-fit parameters for Eq.~\ref{eq:cr_model} were determined separately for the northern and southern hemispheres and for both mono-energetic and kappa-type distributions. These parameters are summarized in Table~\ref{tab:fit_north} for the northern auroral region and Table~\ref{tab:fit_south} for the southern counterpart.

\begin{table}[h]
\centering
\caption{Fit parameters for the CR–energy relationship in the northern hemisphere.}
\label{tab:fit_north}
\begin{tabular}{lcc}
\toprule
\textbf{Parameter} & \textbf{Kappa} & \textbf{Mono-energetic} \\
\midrule
$E_0$ [eV] & 469.86 & 975.38 \\
$A$        & 4.96   & 4.01   \\
$B$ [eV]   & 348426.40 & 624526.56 \\
$C$        & 0.8707 & 0.4707 \\
$D$ [eV]   & 6563.25 & 77356.91 \\
$\alpha$   & 0.9822  & 3.6884 \\
$\beta$    & 2.2396  & 1.5236 \\
$\delta$   & 0.5503  & 1.6034 \\
$\gamma$   & 9.7703  & 10.0226 \\
$\Delta_\text{CR}$ (residual std) & 0.94 & 2.17 \\
\bottomrule
\end{tabular}
\end{table}

\begin{table}[h]
\centering
\caption{Fit parameters for the CR–energy relationship in the southern hemisphere.}
\label{tab:fit_south}
\begin{tabular}{lcc}
\toprule
\textbf{Parameter} & \textbf{Kappa} & \textbf{Mono-energetic} \\
\midrule
$E_0$ [eV] & 346.20 & 1022.80 \\
$A$        & 4.90   & 3.99    \\
$B$ [eV]   & 356080.67 & 544035.84 \\
$C$        & 0.8758 & 0.4145 \\
$D$ [eV]   & 6769.14 & 75619.62 \\
$\alpha$   & 0.9796  & 4.0983 \\
$\beta$    & 2.2662  & 1.4293 \\
$\delta$   & 0.5479  & 1.5626 \\
$\gamma$   & 9.8352  & 10.0418 \\
$\Delta_\text{CR}$ (residual std) & 0.92 & 2.24 \\
\bottomrule
\end{tabular}
\end{table}
The model is valid over the characteristic energy range of approximately 50~eV to 150~keV and across the magnetic latitudes covered by the auroral emissions. 

 \subsection{Flattening of Spectral Profiles Due to Observed High Fluxes}

\begin{figure*}[ht]
    \centering
    
    % ===== Rangée du haut =====
    \begin{subfigure}[b]{0.48\textwidth}
        \centering
        \includegraphics[width=\linewidth]{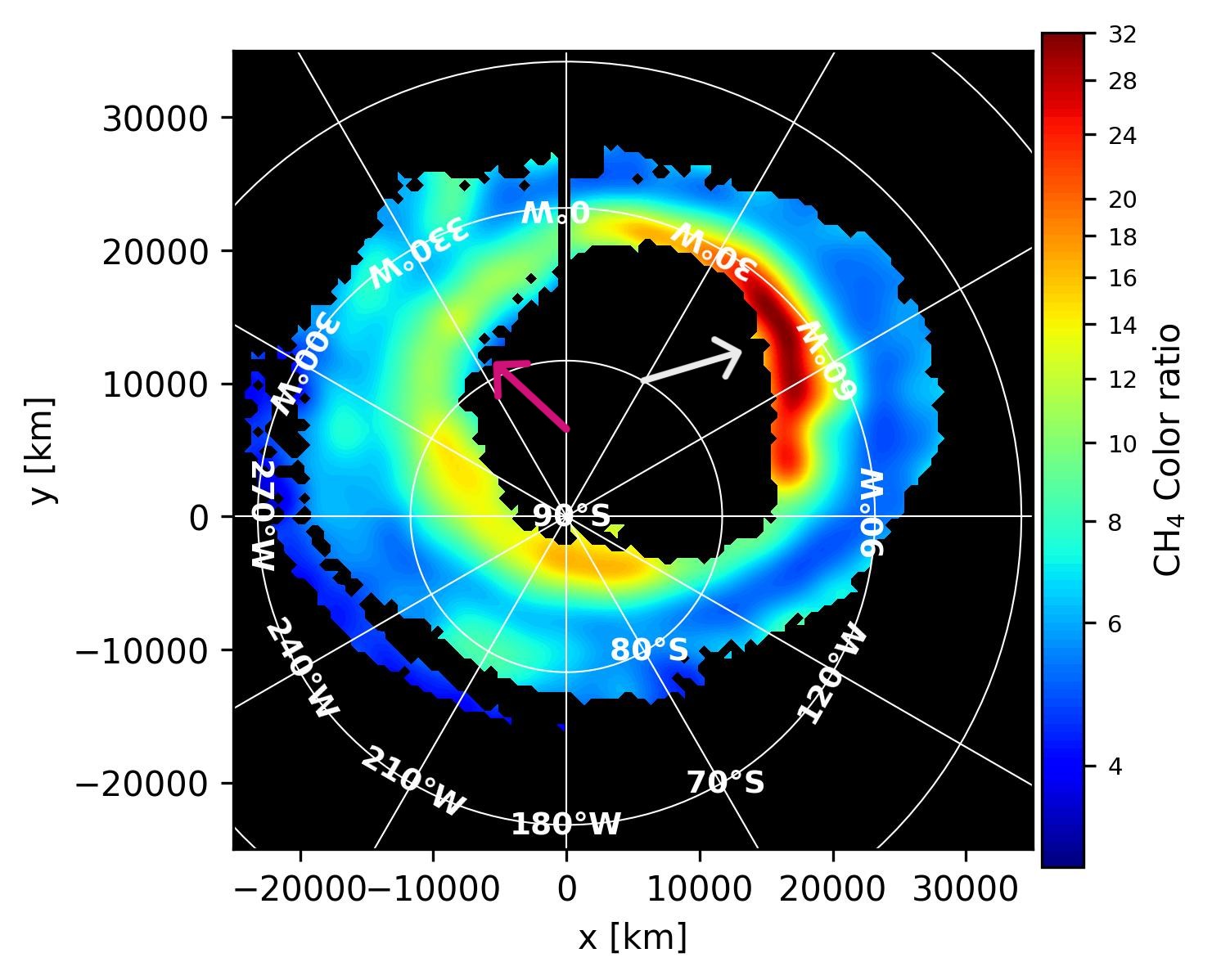}
        \caption{CH$_4$ CR (PJ6).}
        \label{fig:CRCH4_PJ6}
    \end{subfigure}
    \hfill
    \begin{subfigure}[b]{0.48\textwidth}
        \centering
        \includegraphics[width=\linewidth]{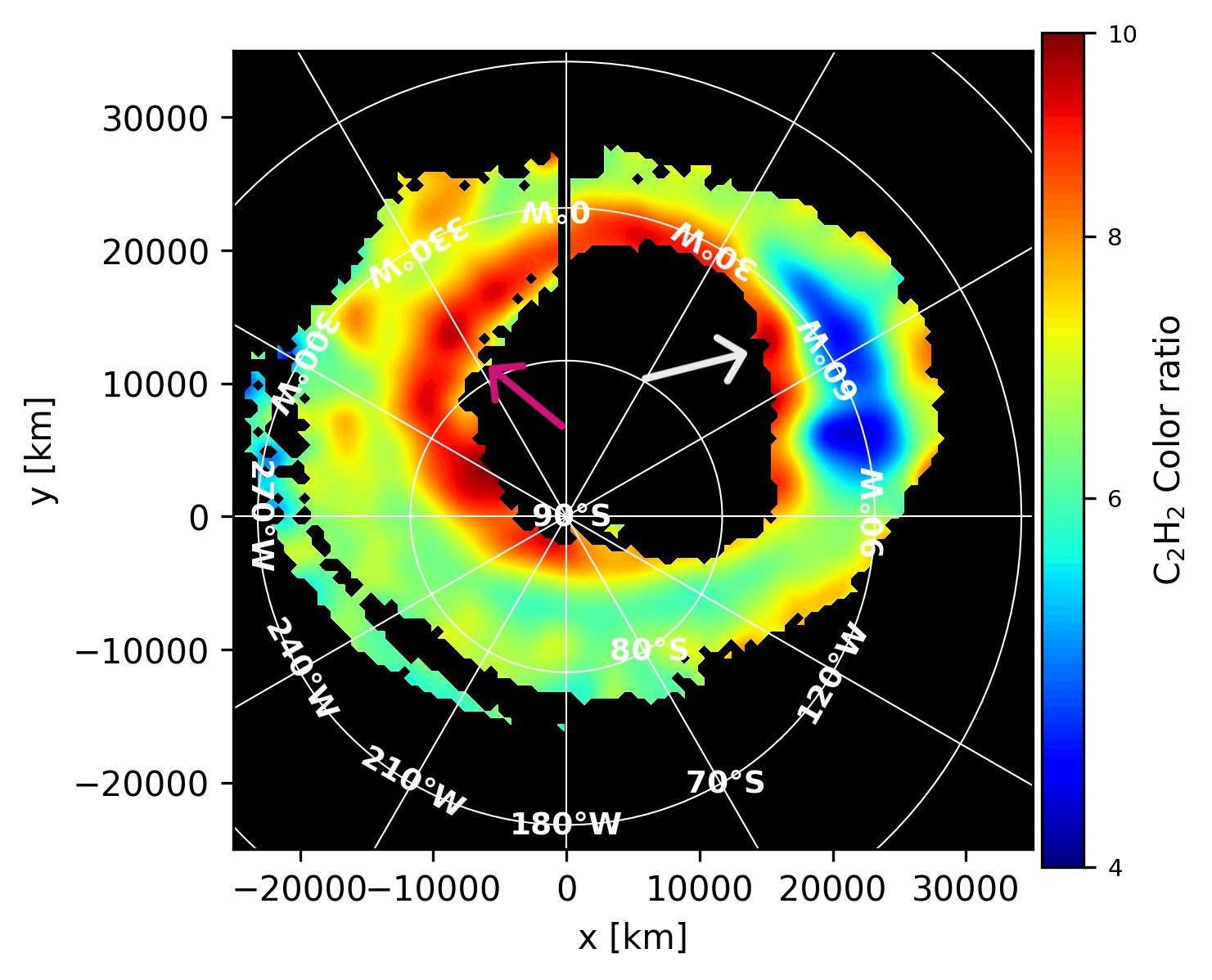}
        \caption{C$_2$H$_2$ CR (PJ6).}
        \label{fig:CRC2H2_PJ6}
    \end{subfigure}
    
    % ===== Rangée du bas =====
    \begin{subfigure}[b]{0.48\textwidth}
        \centering
        \includegraphics[width=\linewidth]{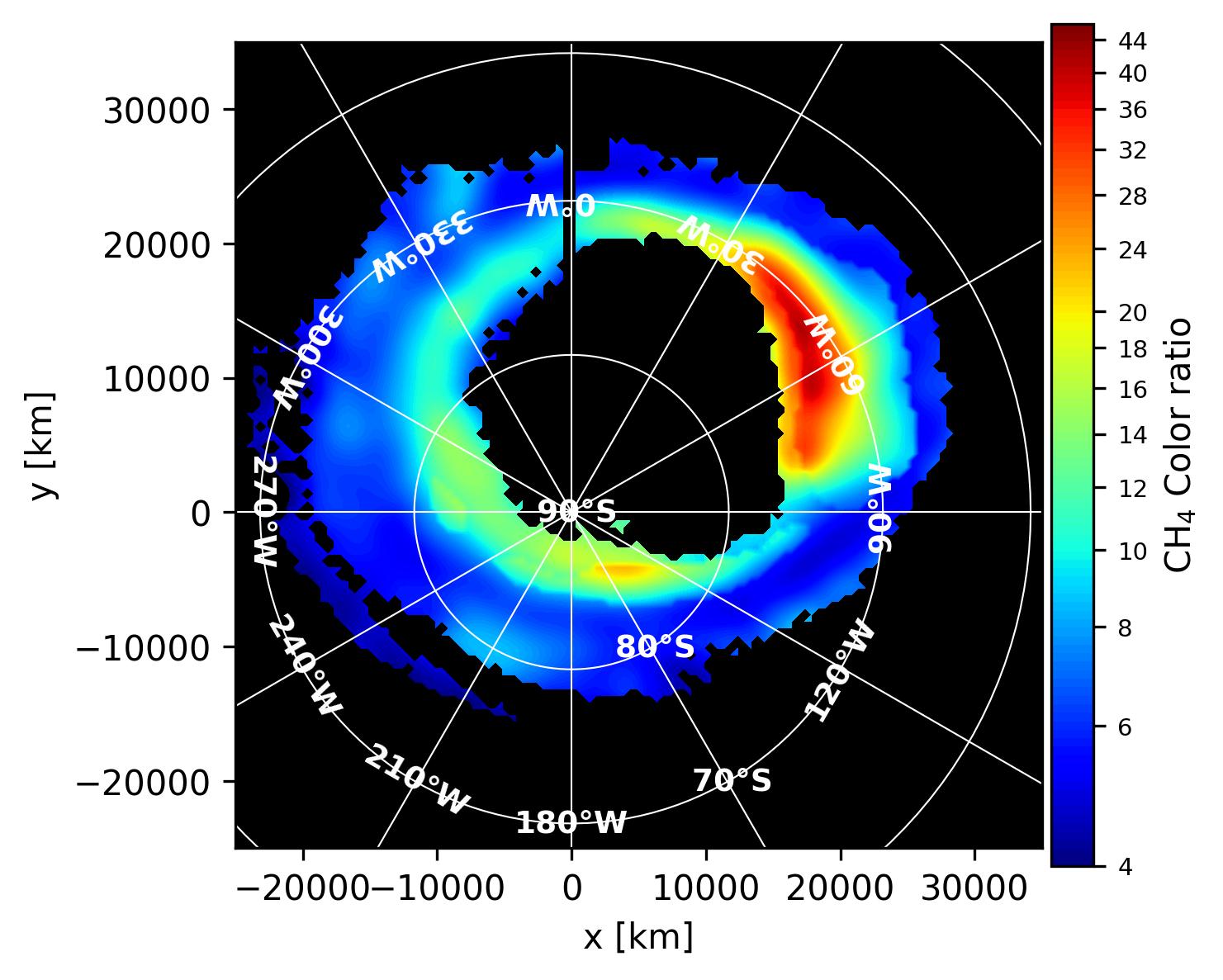}
        \caption{Corrected CH$_4$ CR (PJ6).}
        \label{fig:PJ6_CH4_CR_corrected}
    \end{subfigure}
    \hfill
    \begin{subfigure}[b]{0.48\textwidth}
        \centering
        \includegraphics[width=\linewidth]{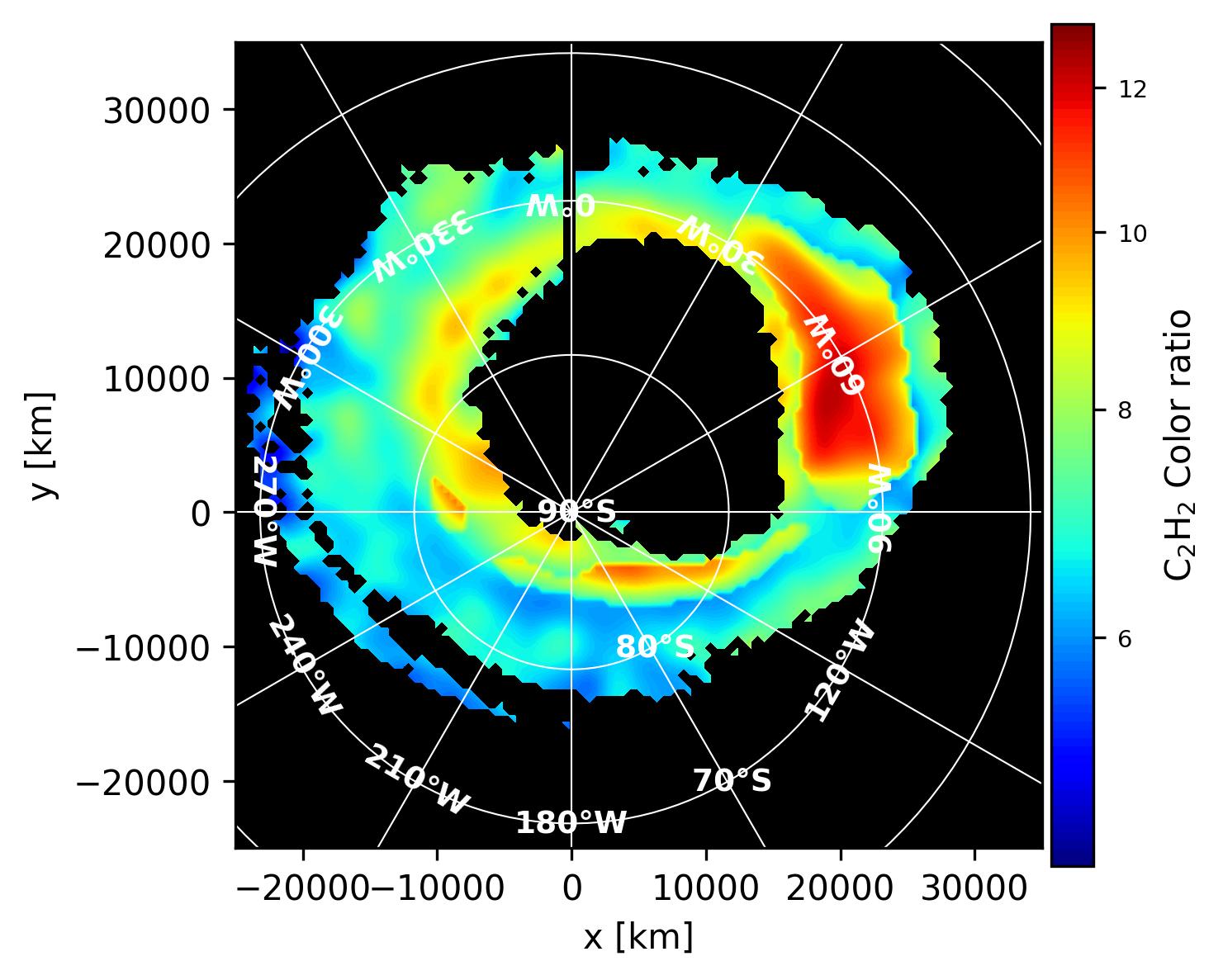}
        \caption{Corrected C$_2$H$_2$ CR (PJ6).}
        \label{fig:PJ6_C2H2_CR_corrected}
    \end{subfigure}
    
    \caption{
        CR maps for PJ6 in the southern auroral region. 
        Panels~\ref{fig:CRCH4_PJ6} and~\ref{fig:CRC2H2_PJ6} show the uncorrected CH$_4$ and C$_2$H$_2$ CR, respectively. 
        A region centered near 70°S, 60°W (white arrow) exhibits strong CH$_4$ absorption but weak C$_2$H$_2$ absorption, 
        while an opposite pattern—low CH$_4$ and enhanced C$_2$H$_2$ absorption—is observed around 75°S, 330°W (pink arrow). 
        Panels~\ref{fig:PJ6_CH4_CR_corrected} and~\ref{fig:PJ6_C2H2_CR_corrected} show the CR maps after applying the spectral flattening correction. 
        The CH$_4$ CR remains largely unchanged, while the C$_2$H$_2$ CR shows stronger absorption in regions previously affected by instrumental effects.
    }
    \label{fig:crmaps_pj6_complete}
\end{figure*}

\begin{figure*}[ht]
    \centering
    
    % ===== Ligne du haut =====
    \begin{subfigure}[b]{0.48\textwidth}
        \centering
        \includegraphics[width=\linewidth]{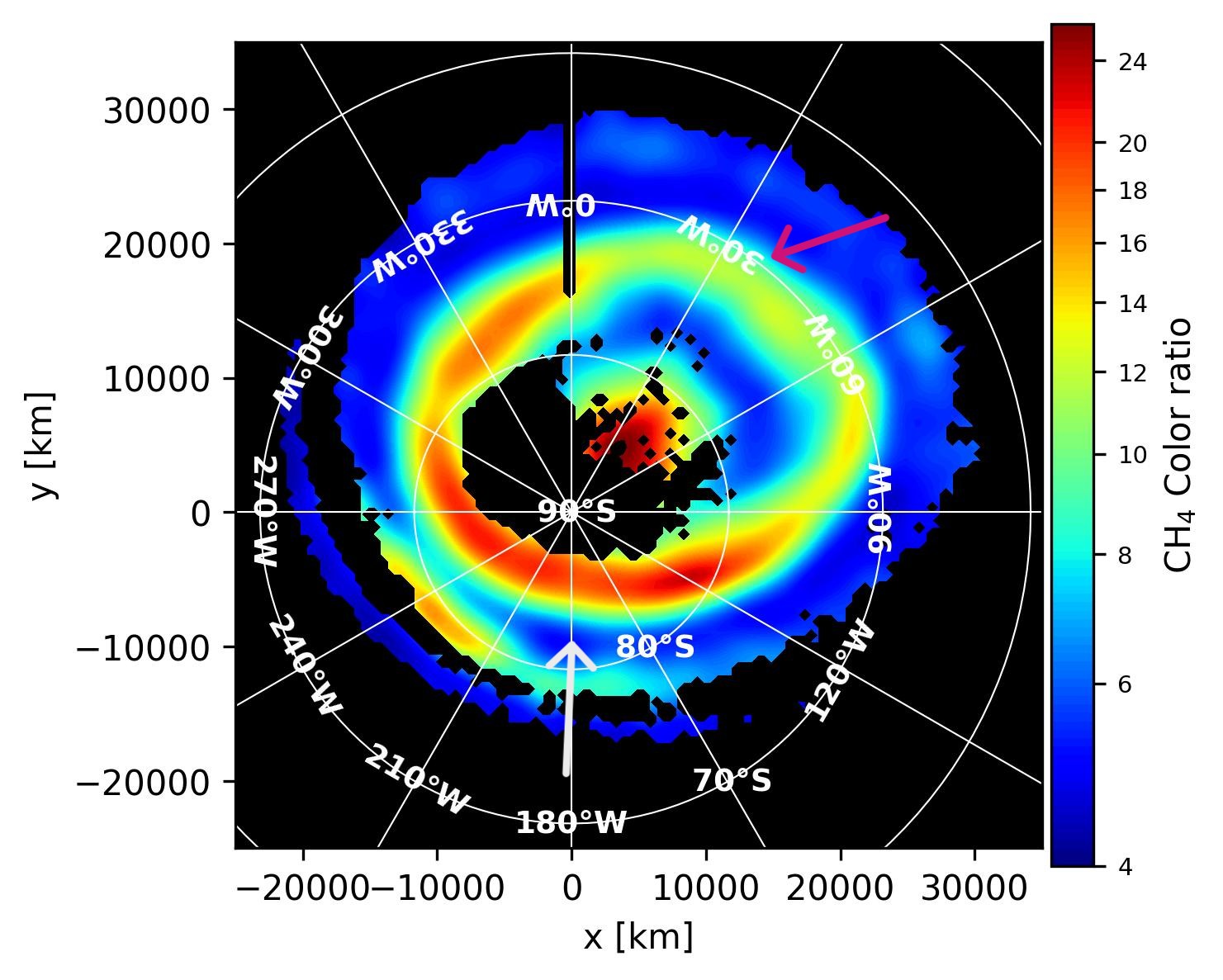}
        \caption{CH$_4$ CR (PJ10).}
        \label{fig:CRCH4_PJ10}
    \end{subfigure}
    \hfill
    \begin{subfigure}[b]{0.48\textwidth}
        \centering
        \includegraphics[width=\linewidth]{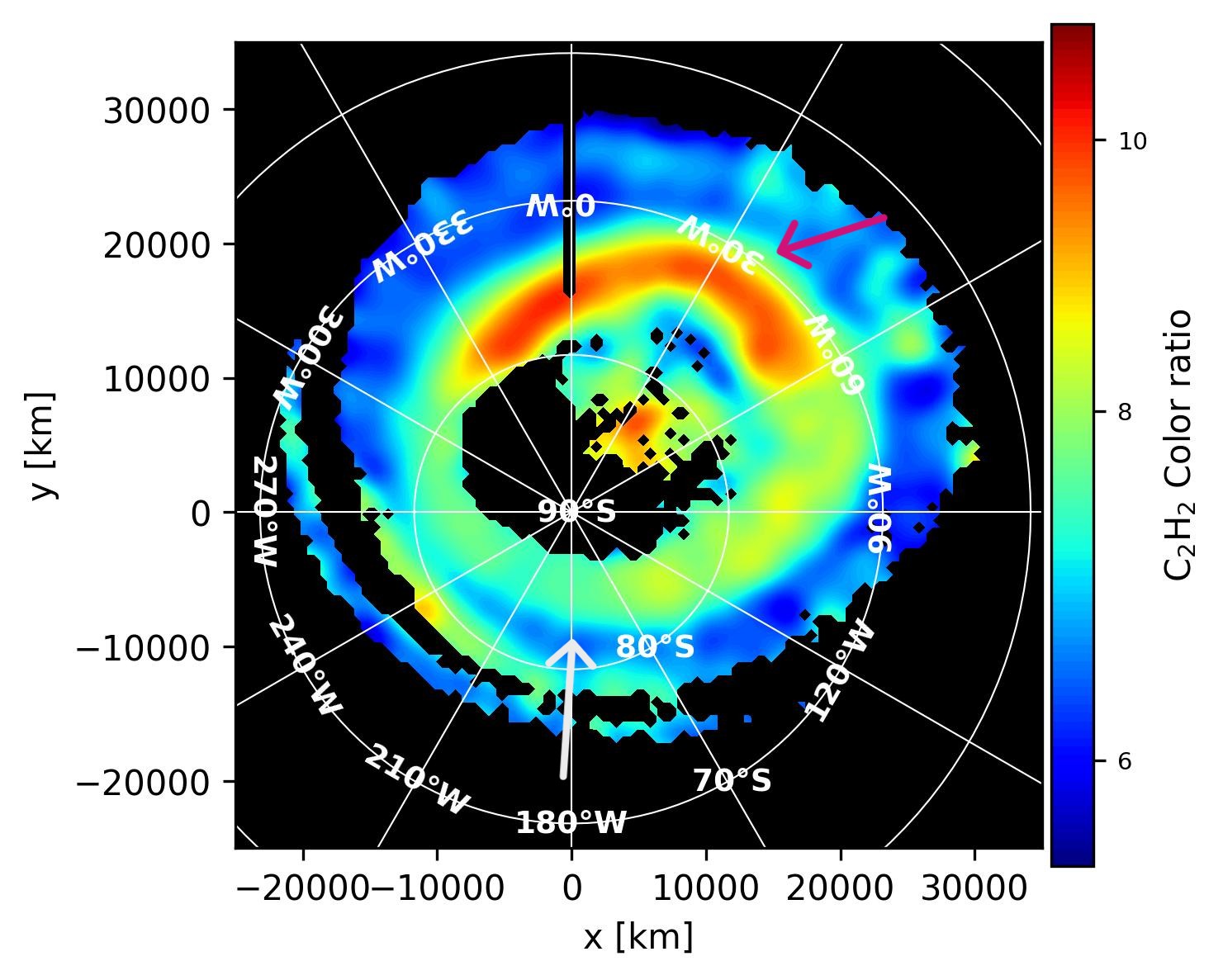}
        \caption{C$_2$H$_2$ CR (PJ10).}
        \label{fig:CRC2H2_PJ10}
    \end{subfigure}
    
    % ===== Ligne du bas =====
    \begin{subfigure}[b]{0.48\textwidth}
        \centering
        \includegraphics[width=\linewidth]{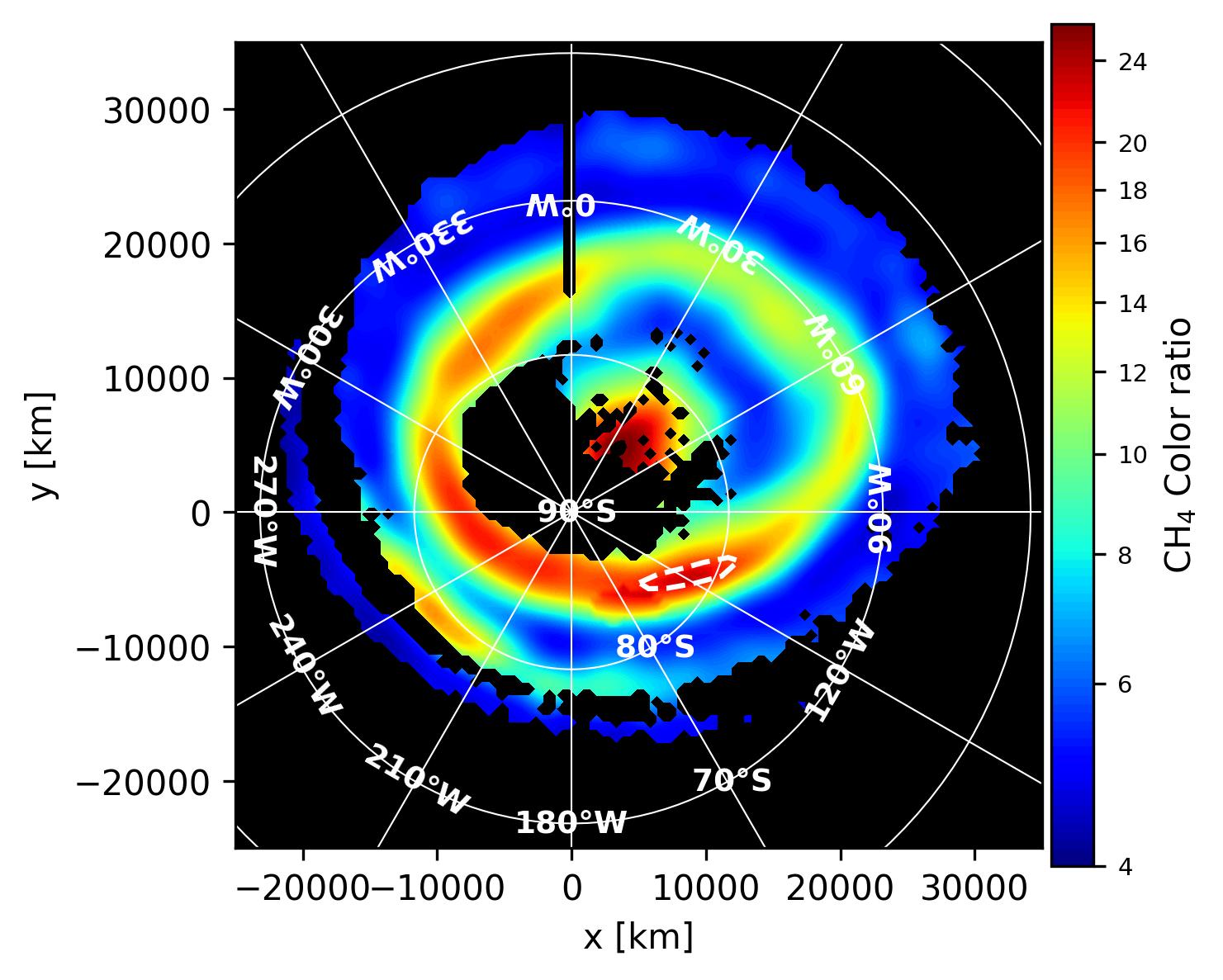}
        \caption{Corrected CH$_4$ CR (PJ10).}
        \label{fig:crmap_ch4_correction_pj10}
    \end{subfigure}
    \hfill
    \begin{subfigure}[b]{0.48\textwidth}
        \centering
        \includegraphics[width=\linewidth]{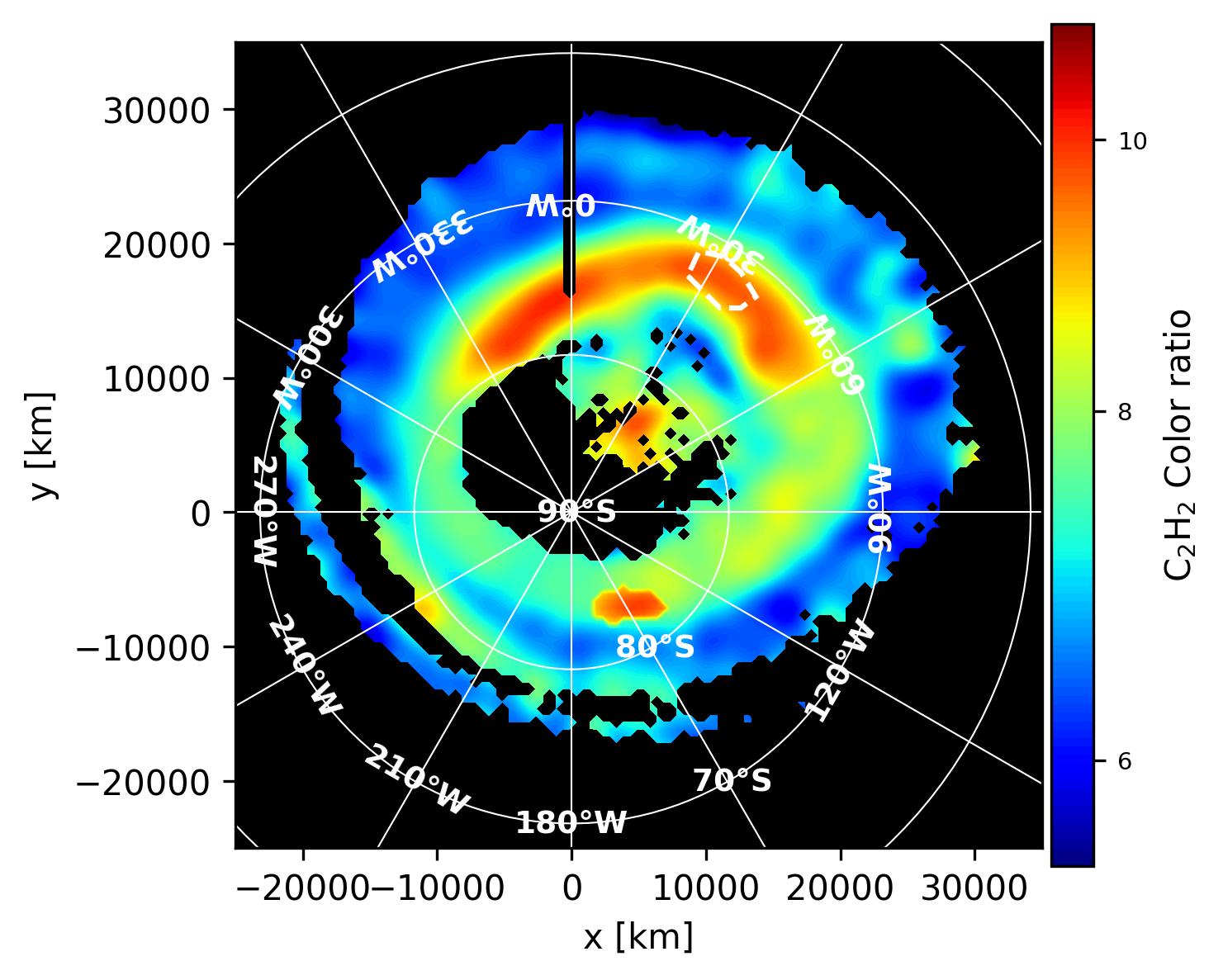}
        \caption{Corrected C$_2$H$_2$ CR (PJ10).}
        \label{fig:crmap_c2h2_correction_pj10}
    \end{subfigure}
    
    \caption{
        CR maps for PJ10 in the southern auroral region. 
        Panels~\ref{fig:CRCH4_PJ10} and~\ref{fig:CRC2H2_PJ10} show the uncorrected CH$_4$ and C$_2$H$_2$ CR, respectively. 
        A region centered near 85°S, 180°W (white arrow) exhibits strong CH$_4$ absorption but weak C$_2$H$_2$ absorption, 
        while an opposite pattern—low CH$_4$ and enhanced C$_2$H$_2$ absorption—is observed around 70°S, 30°W (pink arrow). 
        Panels~\ref{fig:crmap_ch4_correction_pj10} and~\ref{fig:crmap_c2h2_correction_pj10} show the CR maps after applying the spectral flattening correction. 
        The spatial distribution of the CH$_4$ CR remains largely unchanged, indicating that instrumental effects are not the main cause of anomalies. 
        The C$_2$H$_2$ CR shows persistence of anomalous regions, suggesting real atmospheric heterogeneities rather than artifacts. 
        Dashed white polygons outline regions with contrasting CR behaviors.
    }
    \label{fig:crmaps_pj10_complete}
\end{figure*}

Using the revised definition of the CR, based on the 135–140~nm window shown to be unaffected by degradation of the PHD (see Sect.~\ref{sec:cr_definition}), we constructed CH$_4$ and C$_2$H$_2$ CR maps for several perijoves. Figs.~\ref{fig:CRCH4_PJ6}, \ref{fig:CRC2H2_PJ6}, \ref{fig:CRCH4_PJ10} and~\ref{fig:CRC2H2_PJ10} present the results for PJ6 and PJ10 in the southern hemisphere, where unexpected absorption patterns were identified.

Indeed, several regions in both PJ6 and PJ10 exhibit apparent inconsistencies between CH$_4$ and C$_2$H$_2$ absorptions. In each case, we identify two types of  CR anomalies: zones where the CH$_4$ CR is significantly elevated while the C$_2$H$_2$ CR remains low and conversely, zones with high C$_2$H$_2$ CR but unexpectedly weak CH$_4$ CR. These patterns challenge the expected vertical distribution of hydrocarbons and suggest either instrumental effects or local variations in atmospheric composition.

In light of these discrepancies, we also examine the PHD in the 155-162~nm interval, which corresponds to the unabsorbed region of the spectrum and typically exhibits high photon fluxes. In a particularly bright region of PJ6, we find that the PHD peak in this interval is lower than 5 for a particularly bright region corresponding to a dawn storm, suggesting that detector non-linearity effects may also impact this portion of the spectrum and compromise the reliability of the CR in such regions.

\begin{figure}[h]
\centering
\includegraphics[width=0.48\textwidth]{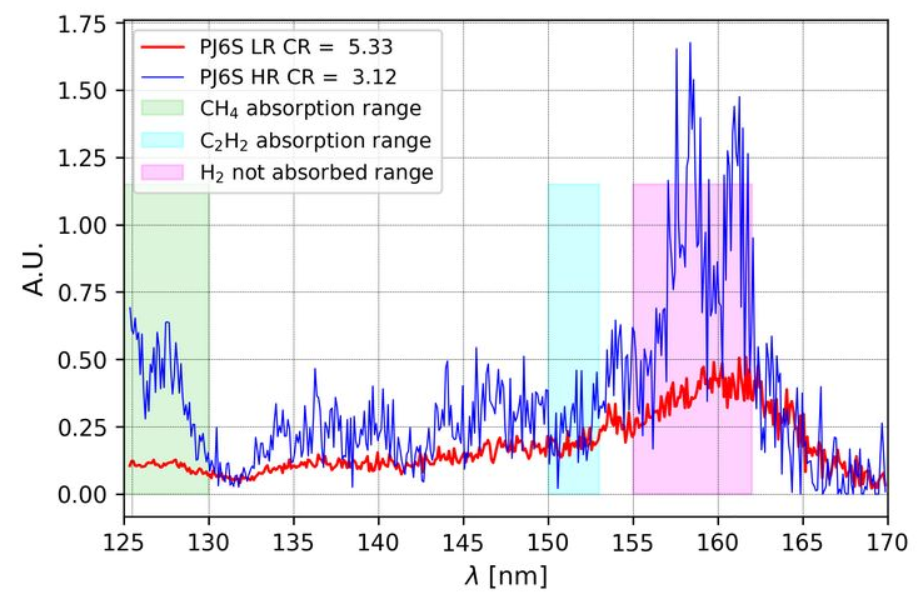}
\caption{Comparison between a spectrum affected by instrumental flattening (red) and an unaffected spectrum (blue), both extracted from the same auroral region during PJ6. The affected spectrum, observed with the wide slit in a very bright area, exhibits a flattened shape with missing emission features particularly in the unabsorbed band. In contrast, the narrow-slit spectrum in the same region clearly shows the characteristic double-peaked H$_2$ continuum around 160 nm.}
\label{fig:pj6_spectrum_anomaly}
\end{figure}

The detailed PHD analysis for the 155--162~nm range is presented in Appendix~\ref{app:erosion}. Affected areas coincide with the brightest portions of the aurora, particularly those associated with a dawn storm, where CR anomalies are most pronounced. In one such region, where the CH$_4$ CR is elevated while the C$_2$H$_2$ CR remains unexpectedly low, we extracted a representative spectrum. As shown in Fig.~\ref{fig:pj6_spectrum_anomaly}, the wide slit observation displays a strong flattening of the H$_2$ emission continuum between 155--162~nm, suppressing the expected double-peaked structure characteristic of this unabsorbed spectral interval. In contrast, the spectrum from the narrow slit retains its expected shape.

These distortions correlate with degraded PHDs in the same region, where the peak falls below a value of 5. This shift toward lower pulse amplitudes is consistent with the detector's high flux non-linearity , likely caused by high photon count rates exceeding the charge recovery capability of the microchannel plate. As a result, photon events are underamplified, leading to gain suppression and spectral flattening , which could be mistakenly ascribed to atmospheric absorption.

We thus expect that in such cases, the apparent suppression of the C$_2$H$_2$ CR and the associated flattening of the spectrum are caused by instrumental effects. To address this, we developed a correction scheme detailed in Appendix~\ref{app:erosion}, which identifies and adjusts CR values in saturated regions, restoring a more reliable diagnostic of hydrocarbon absorption.

To address this, we developed a correction scheme (Appendix~\ref{app:erosion}) that detects and adjusts CR values in regions affected by detector high-flux non-linearity. The method compares the intensities in two narrow diagnostic bands within the unabsorbed part of the H$_2$ spectrum: band A (157–158.5,nm), which contains a sharp emission peak highly sensitive to spectral flattening and band B (163–165,nm), a nearby continuum region largely immune to non-linearity effects. Pixels where the observed A/B ratio falls significantly below the theoretical value from TransPlanet simulations are flagged as degraded and their unabsorbed-band flux is reconstructed from band~B using the theoretical ratio, allowing the CH$_4$ and C$_2$H$_2$ CRs to be corrected accordingly.

The CH$_4$ CR is comparatively less affected by this instrumental effect. The 135–140 nm range lies in a part of the detector that generally receives low local fluxes, both because H$_2$ emissions are intrinsically weaker there and because they are further attenuated by CH$_4$ absorption. As a result, this spectral interval rarely experiences strong non-linear effects and the CH$_4$ CR remains relatively stable even in bright auroral regions. Consequently, the conclusions drawn from CH$_4$ CR maps regarding hydrocarbon absorption largely hold.

In contrast, the 150–153 nm range used for C$_2$H$_2$ absorption lies within a spectral region (150–165 nm) where H$_2$ emissions are much stronger. Here, high local count rates occur across both the absorbed and unabsorbed portions of the spectrum, making the C$_2$H$_2$ CR much more sensitive to detector non-linearities. These effects in the unabsorbed range can then artificially lower the C$_2$H$_2$ CR, even when substantial absorption is present in the 150–153 nm interval. As a result, the corrected CR values provide a more accurate diagnostic of hydrocarbon absorption in high-brightness auroral zones.

\section{Results}
\subsection{Perijove 6}\label{casestudies}

Figures \ref{fig:PJ6_CH4_CR_corrected} and \ref{fig:PJ6_C2H2_CR_corrected} illustrate the outcome of applying the spectral flattening correction to PJ6. The correction restores CR values that are coherent with those predicted by our reference atmospheric model \citep{Grodent2001}. Once corrected, the region initially identified as exhibiting an anomalously low C$_2$H$_2$ CR and an enhanced CH$_4$ CR instead reveals significant absorption in both hydrocarbon bands, consistent with the expected atmospheric structure. This supports the idea that that the anomaly was not atmospheric in nature but instead originated from instrumental effects linked to high flux non-linearity detector effects. 

\subsection{Perijove 10}

The same spectral flattening correction method was applied to PJ10 to assess whether the CR anomalies identified in the initial CR maps could be attributed to detector-induced effects. Pixels were flagged as saturated using the same diagnostic A/B ratio criterion and corrected CR maps were produced. As illustrated in Figs.~\ref{fig:crmap_ch4_correction_pj10} and~\ref{fig:crmap_c2h2_correction_pj10}, these corrected maps exhibit only minor differences compared to the original, uncorrected ones (Figs.~\ref{fig:CRCH4_PJ10} and~\ref{fig:CRC2H2_PJ10}). This indicates that the apparent discrepancies between CH$_4$ and C$_2$H$_2$ absorption features in PJ10 are not instrumental in origin.

Unlike PJ6, where the anomalous horizontal distribution of the CRs was fully resolved after correction, the persistent CR anomalies in PJ10 suggest a physical rather than instrumental cause. These features may reflect real horizontal or vertical variations in hydrocarbon composition, or possibly changes in local atmospheric structure such as temperature, or mixing ratios. This result underscores the necessity of correcting for instrumental effects before interpreting spectral morphologies and points to a potential deviation from the atmospheric model in the auroral regions of PJ10.

\section{Discussion}
The results of this study validate the methodological advances introduced to improve hydrocarbon absorption diagnostics in Juno-UVS data. Redefining the CH$_4$ CR with the 135–140 nm interval yields robust results consistent with radiative transfer models and observational trends in non-saturated regions, while being less affected by instrumental artifacts than the traditional 125–130 nm band. In parallel, the correction method applied to the unabsorbed 155–162 nm range restores reliable CR maps in high-brightness regions in PJ6 by identifying saturated pixels via the A/B band ratio (Appendix~\ref{app:erosion}) and mitigating detector non-linearity effects. This analysis further reveals that the brightness of H$_2$ emission in the 155–162 nm band has been systematically underestimated in regions of highest flux, with the dawn storm during PJ6 showing a deficit of about 10%.

However, despite the success of these corrections in PJ6, significant CR anomalies remain in PJ10. To assess their origin, we select two polygonal regions (see Figs.\ref{fig:crmap_ch4_correction_pj10} and\ref{fig:crmap_c2h2_correction_pj10}) showing contrasting behaviors: one with high CH$_4$ but low C$_2$H$_2$ CR, and another with the opposite pattern. For each polygon, we compute the average spectrum to evaluate whether these anomalies reflect genuine atmospheric structure.

Using the TransPlanet electron transport model, we simulate the FUV auroral spectra resulting from a kappa-distributed population of precipitating electrons passing through the Jovian atmosphere, using the \citet{Grodent2001} atmospheric reference model. The emission angle and mean electron energy for each simulation are fixed based on the median values derived from the data for each polygon and the CR($E_0$) relationship. For both simulations, we allowed for deviations from the reference atmospheric composition by introducing multiplicative adjustment factors ($F_{\mathrm{CH}_4}$, $F_{\mathrm{C}_2\mathrm{H}_2}$, $F_{\mathrm{C}_2\mathrm{H}_6}$) applied to the vertical density profiles of the main hydrocarbons.

\begin{figure}[ht]
    \centering

    \begin{subfigure}[b]{\columnwidth}
        \centering
        \includegraphics[width=\columnwidth]{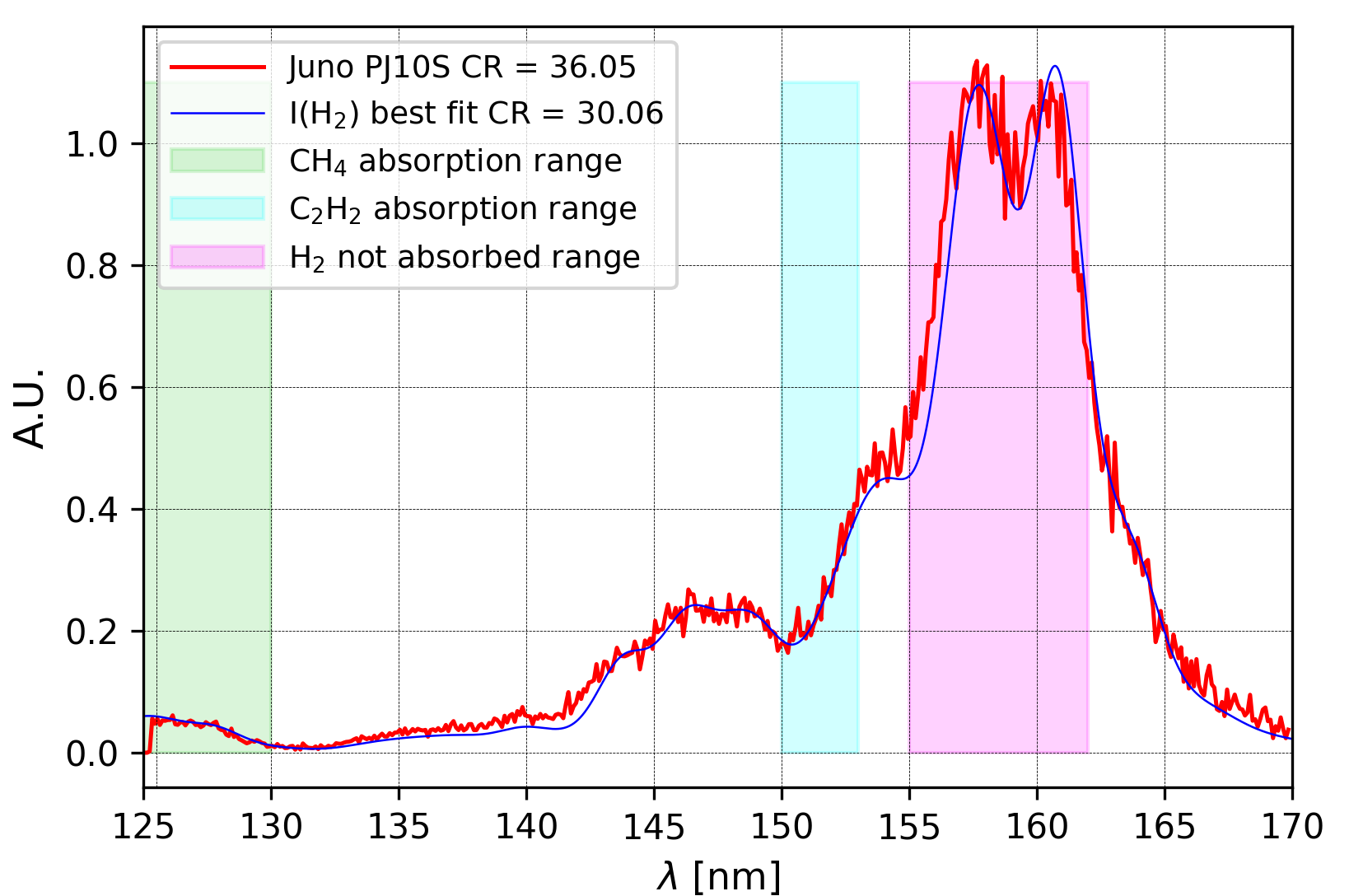}
        \caption{Observed (red) and modeled (blue) FUV spectra for the region shown in Fig.~\ref{fig:crmap_ch4_correction_pj10}. The best fit is obtained with $E_0 = 54.23$~keV and adjustment factors $F_{\mathrm{CH}_4} = 2.0$, $F_{\mathrm{C}_2\mathrm{H}_2} = 0.3$, $F_{\mathrm{C}_2\mathrm{H}_6} = 0.4$. This fit requires a CH$_4$ enhancement and C$_2$H$_2$ depletion, suggesting a local atmospheric anomaly.}
        \label{fig:pj10_fit_polygon1}
    \end{subfigure}

    \vspace{0.5cm}

    \begin{subfigure}[b]{\columnwidth}
        \centering
        \includegraphics[width=\columnwidth]{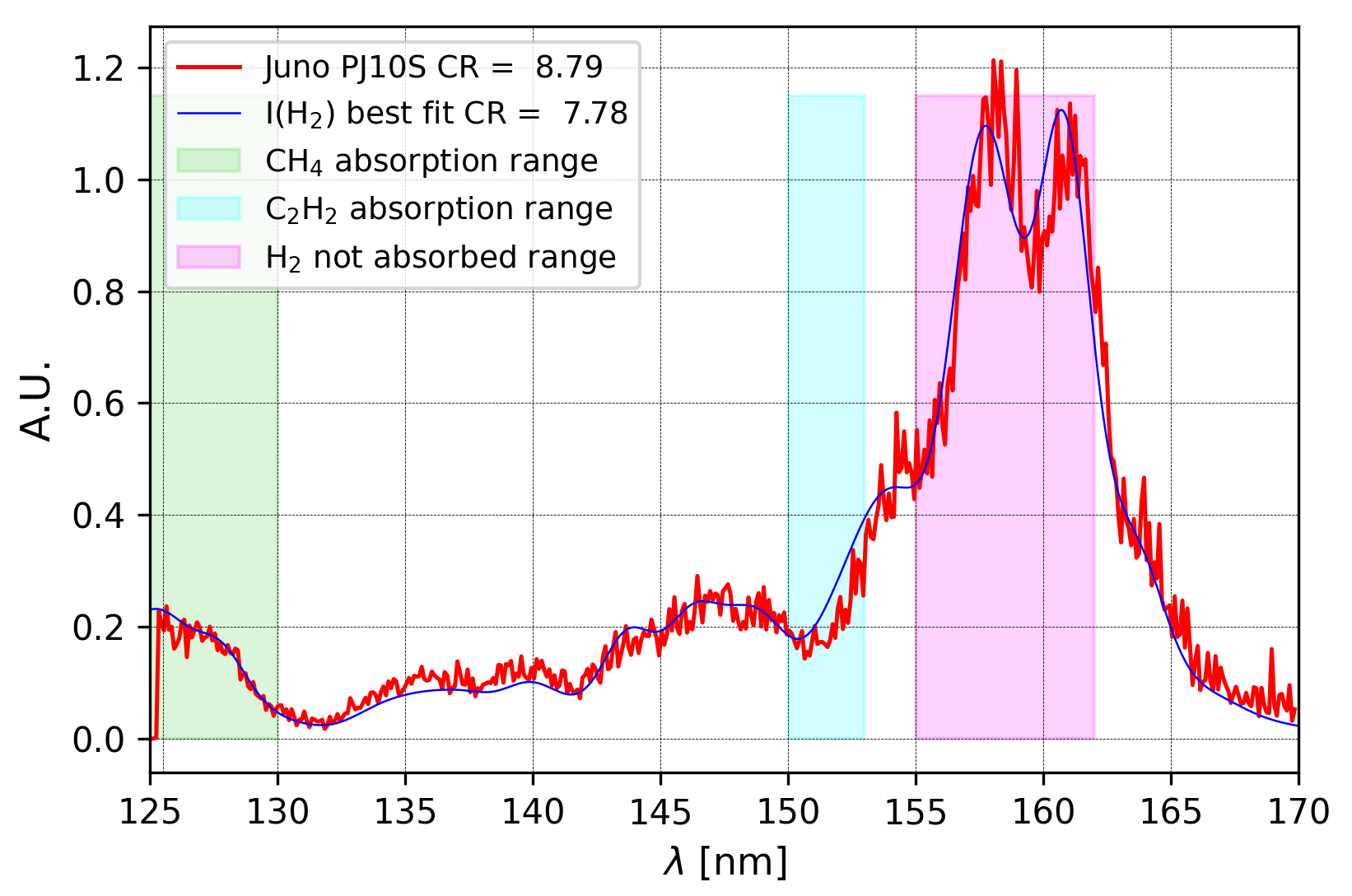}
        \caption{Observed (red) and modeled (blue) spectra for the region in Fig.~\ref{fig:crmap_c2h2_correction_pj10}. The best fit is obtained with $E_0 = 15$~keV and adjustment factors $F_{\mathrm{CH}_4} = 1.0$, $F_{\mathrm{C}_2\mathrm{H}_2} = 1.0$, $F_{\mathrm{C}_2\mathrm{H}_6} = 1.0$. This configuration is consistent with the model atmosphere.}
        \label{fig:pj10_fit_polygon2}
    \end{subfigure}

    \caption{Observed and modeled FUV spectra for two regions in PJ10. The upper panel corresponds to a region with CH$_4$ enhancement and C$_2$H$_2$ depletion, indicative of a local atmospheric anomaly. The lower panel shows a region consistent with the model atmosphere.}
    \label{fig:pj10_fit_comparison}
\end{figure}

In the region with a high CH$_4$ CR and mild C$_2$H$_2$ CR (see Fig.~\ref{fig:crmap_ch4_correction_pj10}), the best fit (see Fig.~\ref{fig:pj10_fit_polygon1}) was obtained with adjustment factors $F_{\mathrm{CH}_4}=2.0$, $F_{\mathrm{C}_2\mathrm{H}_2}=0.3$ and $F_{\mathrm{C}_2\mathrm{H}_6}=0.4$. These values suggest a local enhancement of CH$_4$ and a depletion of C$_2$H$_2$ compared to the reference atmosphere, consistent with the observed CR anomaly. In contrast, for the second region (Fig.~\ref{fig:crmap_c2h2_correction_pj10}), where the CH$_4$ CR is mild and the C$_2$H$_2$ CR is high, the best fit (see Fig.~\ref{fig:pj10_fit_polygon2}) was obtained with $F_{\mathrm{CH}_4}=1.0$, $F_{\mathrm{C}_2\mathrm{H}_2}=1.0$ and $F_{\mathrm{C}_2\mathrm{H}_6}=1.0$, consistent with the nominal composition.

The spectral modeling performed on PJ10 suggests that the anomalous signature observed in the region with high CH$_4$ and mild C$_2$H$_2$ CRs cannot be explained by known instrumental effects such as detector non-linearity at high flux levels. Instead, the best fits require significant deviations from the nominal hydrocarbon profiles, notably an enhanced CH$_4$ abundance and a depleted C$_2$H$_2$ column, incompatible with the standard atmospheric model of \citet{Grodent2001}. These findings point to genuine horizontal heterogeneities and temporal variations in the composition of Jupiter’s upper atmosphere. Importantly, our fitting approach only considered multiplicative scaling factors on the vertical profiles of hydrocarbons.

Our results therefore suggest genuine horizontal heterogeneities in auroral hydrocarbon abundances. This interpretation aligns with previous Juno-UVS solar reflection observations \citep{giles_enhanced_2023} and ground-based mid-infrared measurements \citep{sinclair_2017, sinclair_2018, sinclair_2020, sinclair_2025}, which revealed horizontal variability in CH$_4$ and C$_2$H$_2$ homopauses. 
Unlike these earlier studies, which focused on altitude differences of individual species, our analysis reveals horizontal variations in their relative distributions, an aspect not previously documented.

Possible drivers of this variability include (i) auroral-induced chemistry from electron precipitation, which can locally alter hydrocarbon abundances \citep{sinclair_2019, sinclair_2023, hue2024polar}, and (ii) atmospheric transport through winds or vertical mixing, redistributing species laterally \citep{hue2024polar}. Both processes affect the homopause altitude, which strongly influences UV absorption and thus the observed CRs.

The detection of robust, localized CR anomalies not attributable to instrumental effects demonstrates the diagnostic potential of CR maps beyond energy retrieval. In addition to constraining electron precipitation, they provide insights into atmospheric composition and dynamics, acting as tracers of both energy deposition and the atmospheric response to magnetospheric forcing.

\section{Conclusions}

In this study, we investigated the use of ultraviolet CRs derived from Juno-UVS observations as diagnostics for both electron precipitation energies and atmospheric composition in Jupiter’s auroral regions. By redefining the CH$_4$-sensitive CR to circumvent PHD anomalies in the 125–130~nm interval and by introducing a correction method for detector non-linearity in high-brightness regions, we significantly improved the reliability of the retrieved CR maps.

Applying this methodology to PJ6 and PJ10, we showed that some apparent CR anomalies such as unexpectedly low C$_2$H$_2$ absorption might in fact be artifacts of detector non-linearity at high flux levels. In PJ6, our correction scheme successfully mitigates these effects, restoring physically consistent morphologies. However, in PJ10, the CR anomalies persist even after correction. Spectral fitting with the \texttt{TransPlanet} model suggests that these anomalies reflect true deviations from the reference atmospheric structure, requiring modified CH$_4$ and C$_2$H$_2$ abundances to reproduce the observed spectra.

These results highlight the dual sensitivity of CR diagnostics to particle energy deposition and to atmospheric composition. They also emphasize the limitations of static, one-dimensional models when interpreting auroral spectra, particularly in regions where horizontal gradients or auroral-driven chemistry likely play a role.

The framework developed in this work provides a robust path forward for extracting physical information from ultraviolet auroral emissions. Future efforts will expand this analysis to the full Juno mission dataset, enabling statistical studies of atmospheric variability and offering new constraints on magnetosphere–atmosphere coupling at Jupiter.

\begin{acknowledgements}
      V. Hue \& B. Benmahi acknowledge support from the French government under the France 2030 investment plan, as part of the Initiative d’Excellence d’Aix-Marseille Université – A*MIDEX AMX-22-CPJ-04. French authors acknowledge the support of CNES to the Juno and JUICE missions.
      This work was supported by the Fonds de la Recherche
      Scientifique – FNRS under Grant(s) No. T003524F. B. Bonfond is a Research Associate of the Fonds de la Recherche Scientifique - FNRS.
\end{acknowledgements}


\bibliographystyle{aa} % style aa.bst
\bibliography{These.bib} % your references Yourfile.bib

\begin{thebibliography}{48}
\expandafter\ifx\csname natexlab\endcsname\relax\def\natexlab#1{#1}\fi

\bibitem[{Acton {et~al.}(2018)Acton, Bachman, Semenov, \& Wright}]{ACTON20189}
Acton, C., Bachman, N., Semenov, B., \& Wright, E. 2018, Planetary and Space Science, 150, 9, enabling Open and Interoperable Access to Planetary Science and Heliophysics Databases and Tools

\bibitem[{Acton(1996)}]{ACTON199665}
Acton, C.~H. 1996, Planetary and Space Science, 44, 65, planetary data system

\bibitem[{Ajello {et~al.}(2005)Ajello, Pryor, Esposito, Stewart, McClintock, Gustin, Grodent, G{\'e}rard, \& Clarke}]{ajello_properties_2005}
Ajello, J.~M., Pryor, W., Esposito, L., {et~al.} 2005, Icarus, 178, 327

\bibitem[{Bagenal {et~al.}(2017)Bagenal, Adriani, Allegrini, Bolton, Bonfond, Bunce, Connerney, Cowley, Ebert, Gladstone, {et~al.}}]{Bagenal2017}
Bagenal, F., Adriani, A., Allegrini, F., {et~al.} 2017, Space Science Reviews, 213, 219

\bibitem[{Benmahi(2022)}]{benmahi_etude_2022}
Benmahi, B. 2022, Theses, {Universit{\'e} de Bordeaux}

\bibitem[{Benmahi {et~al.}(2024{\natexlab{a}})Benmahi, Bonfond, Benne, Grodent, Hue, Gladstone, Gronoff, Lilensten, Sicorello, Head, Barthélemy, Wedlund, Giles, \& Greathouse}]{benmahi_energy_2024}
Benmahi, B., Bonfond, B., Benne, B., {et~al.} 2024{\natexlab{a}}, A\&A, 685, A26, publisher: EDP Sciences

\bibitem[{Benmahi {et~al.}(2024{\natexlab{b}})Benmahi, Bonfond, Benne, Hue, Grodent, Barthélemy, Sinclair, Moirano, Head, Gladstone, Gronoff, Sicorello, Wedlund, Giles, \& Greathouse}]{benmahi_auroral_2024}
Benmahi, B., Bonfond, B., Benne, B., {et~al.} 2024{\natexlab{b}}, A\&A, 691, A91, publisher: EDP Sciences

\bibitem[{Benmahi {et~al.}(2020)Benmahi, Cavalié, Dobrijevic, Biver, Bermudez~Diaz, Sandqvist, Lellouch, Moreno, Fouchet, Hue, Hartogh, Billebaud, Lecacheux, Hjalmarson, Frisk, \& Olberg}]{benmahi_monitoring_2020}
Benmahi, B., Cavalié, T., Dobrijevic, M., {et~al.} 2020, EPSC2020, conference Name: European Planetary Science Congress ADS Bibcode: 2020EPSC...14...87B

\bibitem[{Benne {et~al.}(2024)Benne, Benmahi, Dobrijevic, Cavalié, Loison, Hickson, Barthélémy, \& Lilensten}]{benne_impact_2024}
Benne, B., Benmahi, B., Dobrijevic, M., {et~al.} 2024, A\&A, 686, A22, publisher: EDP Sciences

\bibitem[{Bolton {et~al.}(2017)Bolton, Adriani, Adumitroaie, Allison, Anderson, Atreya, Bloxham, Brown, Connerney, DeJong, {et~al.}}]{Bolton2017}
Bolton, S.~J., Adriani, A., Adumitroaie, V., {et~al.} 2017, Science, 356, 821

\bibitem[{Bonfond {et~al.}(2017)Bonfond, Gladstone, Grodent, Greathouse, Versteeg, Hue, Davis, Vogt, G{\'e}rard, Radioti, {et~al.}}]{bonfond2017morphology}
Bonfond, B., Gladstone, G., Grodent, D., {et~al.} 2017, Geophysical Research Letters, 44, 4463

\bibitem[{Broadfoot {et~al.}(1979)Broadfoot, Belton, Takacs, Sandel, Shemansky, Holberg, Ajello, Atreya, Donahue, Moos, Bertaux, Blamont, Strobel, McConnell, Dalgarno, Goody, \& McElroy}]{broadfoot_extreme_1979}
Broadfoot, A.~L., Belton, M. J.~S., Takacs, P.~Z., {et~al.} 1979, Science, publisher: American Association for the Advancement of Science

\bibitem[{Cavalié {et~al.}(2023)Cavalié, Rezac, Moreno, Lellouch, Fouchet, Benmahi, Greathouse, Sinclair, Hue, Hartogh, Dobrijevic, Carrasco, \& Perrin}]{cavalie_evidence_2023}
Cavalié, T., Rezac, L., Moreno, R., {et~al.} 2023, Nature Astronomy, 7, 1048, aDS Bibcode: 2023NatAs...7.1048C

\bibitem[{Clarke {et~al.}(1980)Clarke, Moos, Atreya, \& Lane}]{Clarke1980}
Clarke, J., Moos, H., Atreya, S., \& Lane, A. 1980, Astrophysical Journal, Part 2-Letters to the Editor, vol. 241, Nov. 1, 1980, p. L179-L182., 241, L179

\bibitem[{Clarke {et~al.}(1998)Clarke, Ballester, Trauger, Ajello, Pryor, Tobiska, Connerney, Gladstone, Waite~Jr, Ben~Jaffel, {et~al.}}]{Clarke1998}
Clarke, J.~T., Ballester, G., Trauger, J., {et~al.} 1998, Journal of Geophysical Research: Planets, 103, 20217

\bibitem[{Clarke {et~al.}(1996)Clarke, Ballester, Trauger, Evans, Connerney, Stapelfeldt, Crisp, Feldman, Burrows, Casertano, {et~al.}}]{Clarke1996}
Clarke, J.~T., Ballester, G.~E., Trauger, J., {et~al.} 1996, Science, 274, 404

\bibitem[{Dols {et~al.}(1992)Dols, G{\'e}rard, Paresce, Prang{\'e}, \& Vidal-Madjar}]{Dols1992}
Dols, V., G{\'e}rard, J.-C., Paresce, F., Prang{\'e}, R., \& Vidal-Madjar, A. 1992, Geophysical research letters, 19, 1803

\bibitem[{Foreman-Mackey {et~al.}(2013)Foreman-Mackey, Hogg, Lang, \& Goodman}]{foreman2013emcee}
Foreman-Mackey, D., Hogg, D.~W., Lang, D., \& Goodman, J. 2013, Publications of the Astronomical Society of the Pacific, 125, 306

\bibitem[{G{\'e}rard {et~al.}(1994)G{\'e}rard, Grodent, Dols, Prang{\'e}, Waite, Gladstone, Franke, Paresce, Storrs, \& Jaffel}]{Gerard1994}
G{\'e}rard, J.-C., Grodent, D., Dols, V., {et~al.} 1994, Science, 266, 1675

\bibitem[{Giles {et~al.}(2023)Giles, Hue, Greathouse, Gladstone, Kammer, Versteeg, Bonfond, Grodent, Gérard, Sinclair, Bolton, \& Levin}]{giles_enhanced_2023}
Giles, R.~S., Hue, V., Greathouse, T.~K., {et~al.} 2023, Journal of Geophysical Research (Planets), 128, e2022JE007610, aDS Bibcode: 2023JGRE..12807610G

\bibitem[{Gladstone \& Skinner(1989)}]{Gladstone1989}
Gladstone, G. \& Skinner, T. 1989, NASA SPECIAL PUBLICATION SERIES, NASA-SP-494, 221-228. Document Section., 494

\bibitem[{Gladstone {et~al.}(2017)Gladstone, Versteeg, Greathouse, Hue, Davis, G{\'e}rard, Grodent, Bonfond, Nichols, Wilson, {et~al.}}]{Gladstone2017}
Gladstone, G., Versteeg, M., Greathouse, T., {et~al.} 2017, Geophysical Research Letters, 44, 7668

\bibitem[{Greathouse {et~al.}(2013)Greathouse, Gladstone, Davis, Slater, Versteeg, Persson, Walther, Winters, Persyn, \& Eterno}]{greathouse_performance}
Greathouse, T.~K., Gladstone, G.~R., Davis, M.~W., {et~al.} 2013, in UV, X-Ray, and Gamma-Ray Space Instrumentation for Astronomy XVIII, ed. O.~H. Siegmund, Vol. 8859, International Society for Optics and Photonics (SPIE), 88590T

\bibitem[{Grodent(2015)}]{grodent_brief_2015}
Grodent, D. 2015, Space Science Reviews, 187, 23, aDS Bibcode: 2015SSRv..187...23G

\bibitem[{Grodent {et~al.}(1997)Grodent, Gladstone, G{\'e}rard, Dols, \& Waite}]{Grodent1997}
Grodent, D., Gladstone, G., G{\'e}rard, J.-C., Dols, V., \& Waite, J. 1997, Icarus, 128, 306

\bibitem[{Grodent {et~al.}(2001)Grodent, Waite~Jr, \& G{\'e}rard}]{Grodent2001}
Grodent, D., Waite~Jr, J.~H., \& G{\'e}rard, J.-C. 2001, Journal of Geophysical Research: Space Physics, 106, 12933

\bibitem[{Gustin {et~al.}(2002)Gustin, Grodent, G{\'e}rard, \& Clarke}]{Gustin2002}
Gustin, J., Grodent, D., G{\'e}rard, J.-C., \& Clarke, J. 2002, Icarus, 157, 91

\bibitem[{Gustin {et~al.}(2016)Gustin, Grodent, Ray, Bonfond, Bunce, Nichols, \& Ozak}]{gustin_characterization_2016}
Gustin, J., Grodent, D., Ray, L., {et~al.} 2016, Icarus, 268, 215

\bibitem[{Gérard {et~al.}(2014)Gérard, Bonfond, Grodent, Radioti, Clarke, Gladstone, Waite, Bisikalo, \& Shematovich}]{gerard_mapping_2014}
Gérard, J.-C., Bonfond, B., Grodent, D., {et~al.} 2014, Journal of Geophysical Research: Space Physics, 119, 9072, \_eprint: https://onlinelibrary.wiley.com/doi/pdf/10.1002/2014JA020514

\bibitem[{Gérard {et~al.}(2019)Gérard, Bonfond, Mauk, Gladstone, Yao, Greathouse, Hue, Grodent, Gkouvelis, Kammer, Versteeg, Clark, Radioti, Connerney, Bolton, \& Levin}]{gerard_contemporaneous_2019}
Gérard, J.-C., Bonfond, B., Mauk, B.~H., {et~al.} 2019, Journal of Geophysical Research: Space Physics, 124, 8298, \_eprint: https://onlinelibrary.wiley.com/doi/pdf/10.1029/2019JA026862

\bibitem[{Harris {et~al.}(1996)Harris, Clarke, McGrath, \& Ballester}]{Harris1996}
Harris, W., Clarke, J.~T., McGrath, M.~A., \& Ballester, G.~E. 1996, Icarus, 123, 350

\bibitem[{Hue {et~al.}(2024)Hue, Cavali{\'e}, Sinclair, Zhang, Benmahi, Rodr{\'\i}guez-Ovalle, Giles, Stallard, Johnson, Dobrijevic, {et~al.}}]{hue2024polar}
Hue, V., Cavali{\'e}, T., Sinclair, J.~A., {et~al.} 2024, Space Science Reviews, 220, 85

\bibitem[{Hue {et~al.}(2021)Hue, Giles, Gladstone, Greathouse, Davis, Kammer, \& Versteeg}]{hue2021updated}
Hue, V., Giles, R.~S., Gladstone, G.~R., {et~al.} 2021, Journal of Astronomical Telescopes, Instruments, and Systems, 7, 044003

\bibitem[{Hue {et~al.}(2018)Hue, Kammer, Gladstone, Greathouse, Davis, Bonfond, Versteeg, Grodent, Gérard, Bolton, \& Levin}]{hue_-flight_2018}
Hue, V., Kammer, J.~A., Gladstone, G.~R., {et~al.} 2018, in Space {Telescopes} and {Instrumentation} 2018: {Ultraviolet} to {Gamma} {Ray}, ed. J.-W.~A. Den~Herder, K.~Nakazawa, \& S.~Nikzad (Austin, United States: SPIE), 108

\bibitem[{Lilensten {et~al.}(1989)Lilensten, Kofman, Wisemberg, Oran, \& DeVore}]{lilensten_ionization_1989}
Lilensten, J., Kofman, W., Wisemberg, J., Oran, E.~S., \& DeVore, C.~R. 1989, Annales Geophysicae, 7, 83, publisher: Springer ADS Bibcode: 1989AnGeo...7...83L

\bibitem[{Liu {et~al.}(1995)Liu, Ahmed, Multari, James, \& Ajello}]{Liu1995}
Liu, X., Ahmed, S.~M., Multari, R.~A., James, G.~K., \& Ajello, J.~M. 1995, Astrophysical Journal Supplement v. 101, p. 375, 101, 375

\bibitem[{Livengood(1992)}]{Livengood1992}
Livengood, T.~A. 1992, The Jovian ultraviolet aurora observed with the IUE spacecraft: Brightness and color distribution with longitude (The Johns Hopkins University)

\bibitem[{O’Donoghue {et~al.}(2021)O’Donoghue, Moore, Bhakyapaibul, Melin, Stallard, Connerney, \& Tao}]{o2021global}
O’Donoghue, J., Moore, L., Bhakyapaibul, T., {et~al.} 2021, Nature, 596, 54

\bibitem[{Prang{\'e} {et~al.}(1998)Prang{\'e}, Rego, Pallier, Connerney, Zarka, \& Queinnec}]{Prange1998}
Prang{\'e}, R., Rego, D., Pallier, L., {et~al.} 1998, Journal of Geophysical Research: Planets, 103, 20195

\bibitem[{Sinclair {et~al.}(2020)Sinclair, Greathouse, Giles, Antuñano, Moses, Fouchet, Bézard, Tao, Martín-Torres, Clark, Grodent, Orton, Hue, Fletcher, \& Irwin}]{sinclair_2020}
Sinclair, J.~A., Greathouse, T.~K., Giles, R.~S., {et~al.} 2020, The Planetary Science Journal, 1, 85, publisher: IOP Publishing

\bibitem[{Sinclair {et~al.}(2025)Sinclair, Greathouse, Giles, Richter, Rashman, Witt, Moses, Hue, Rodríguez-Ovalle, Fouchet, Bhattacharya, Benmahi, Orton, Fletcher, \& Irwin}]{sinclair_2025}
Sinclair, J.~A., Greathouse, T.~K., Giles, R.~S., {et~al.} 2025, The Planetary Science Journal, 6, 15, publisher: IOP Publishing

\bibitem[{Sinclair {et~al.}(2019)Sinclair, Orton, Fernandes, Kasaba, Sato, Fujiyoshi, Tao, Vogt, Grodent, Bonfond, Moses, Greathouse, Dunn, Giles, Tabataba-Vakili, Fletcher, \& Irwin}]{sinclair_2019}
Sinclair, J.~A., Orton, G.~S., Fernandes, J., {et~al.} 2019, Nature Astronomy, 3, 607, publisher: Nature Publishing Group

\bibitem[{Sinclair {et~al.}(2017)Sinclair, Orton, Greathouse, Fletcher, Moses, Hue, \& Irwin}]{sinclair_2017}
Sinclair, J.~A., Orton, G.~S., Greathouse, T.~K., {et~al.} 2017, Icarus, 292, 182

\bibitem[{Sinclair {et~al.}(2018)Sinclair, Orton, Greathouse, Fletcher, Moses, Hue, \& Irwin}]{sinclair_2018}
Sinclair, J.~A., Orton, G.~S., Greathouse, T.~K., {et~al.} 2018, Icarus, 300, 305

\bibitem[{Sinclair {et~al.}(2023)Sinclair, West, Barbara, Tao, Orton, Greathouse, Giles, Grodent, Fletcher, \& Irwin}]{sinclair_2023}
Sinclair, J.~A., West, R., Barbara, J.~M., {et~al.} 2023, Icarus, 406, 115740

\bibitem[{Tao {et~al.}(2016)Tao, Kimura, Badman, Andr{\'e}, Tsuchiya, Murakami, Yoshioka, Yoshikawa, Yamazaki, \& Fujimoto}]{tao_variation_2016}
Tao, C., Kimura, T., Badman, S.~V., {et~al.} 2016, Journal of Geophysical Research: Space Physics, 121, 4055

\bibitem[{Trafton {et~al.}(1994)Trafton, Gerard, Munhoven, \& Waite~Jr}]{trafton_fuv_1994}
Trafton, L., Gerard, J.-C., Munhoven, G., \& Waite~Jr, J. 1994, Astrophysical Journal, Part 1 (ISSN 0004-637X), vol. 421, no. 2, p. 816-827, 421, 816

\bibitem[{Yung {et~al.}(1982)Yung, Gladstone, Chang, Ajello, \& Srivastava}]{yung_h2_1982}
Yung, Y.~L., Gladstone, G.~R., Chang, K.~M., Ajello, J.~M., \& Srivastava, S.~K. 1982, The Astrophysical Journal, 254, L65, publisher: IOP ADS Bibcode: 1982ApJ...254L..65Y

\end{thebibliography}

\appendix
\section{Justification for the Redefinition of the CH$_4$ Absorption Band}
\label{app:crband}

The standard CH$_4$ CR uses the 125-130~nm interval as the absorption band \citep[e.g.,][]{gerard_contemporaneous_2019}, due to its strong sensitivity to CH$_4$. However, a detailed analysis of the Juno-UVS data revealed that this interval is frequently affected by detector calibration issues, notably due to distortions in the PHD. These artifacts primarily affect the wide slit measurements and manifest as underestimation of photon counts in high-brightness regions, leading to unreliable spectral intensities.

\begin{figure}[h]
\centering
\includegraphics[width=0.48\textwidth]{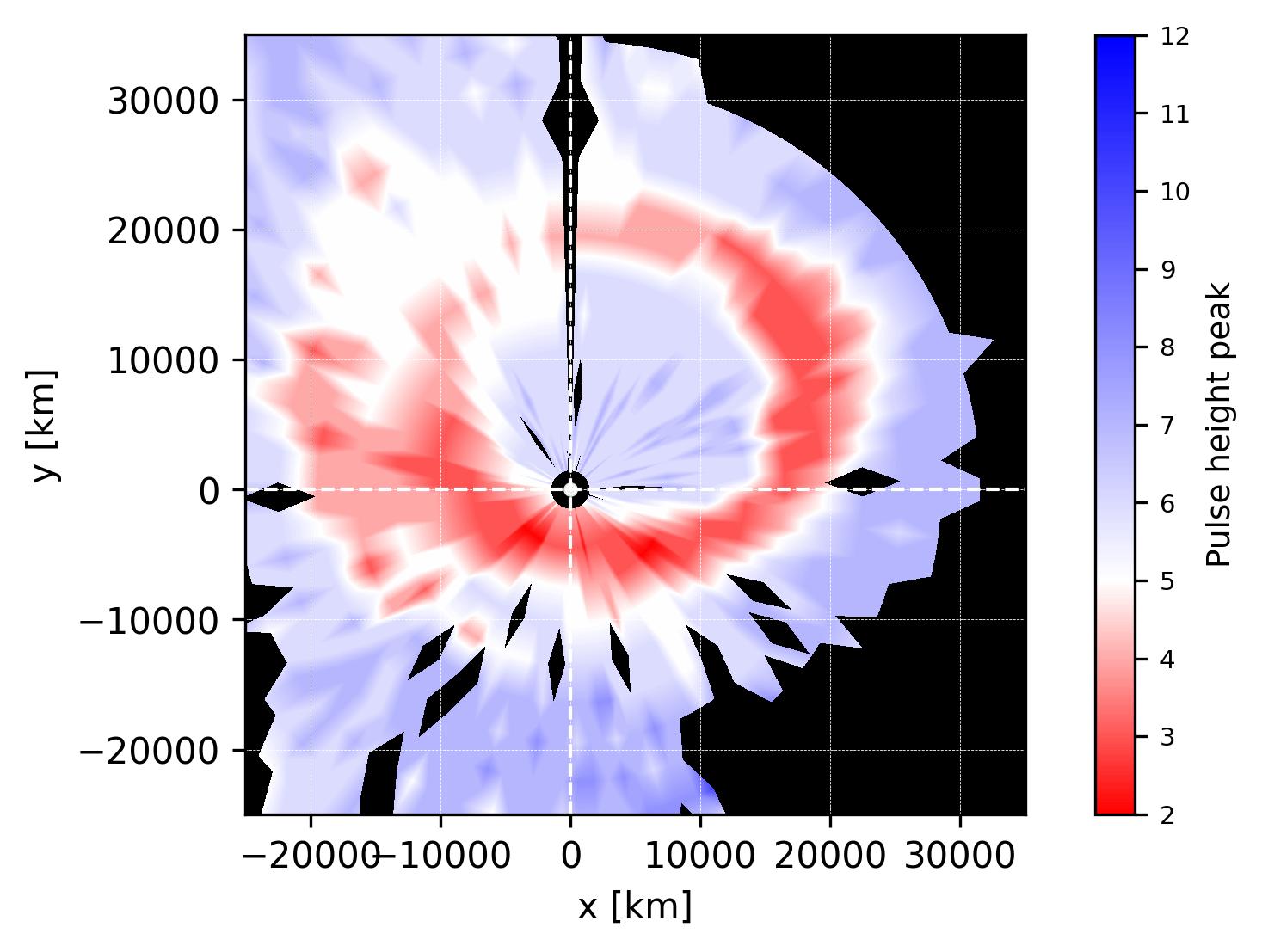}
\caption{Pulse height distribution (PHD) peak map for the 125-130 nm interval during the whole duration of PJ6. Regions where the PHD peak falls below the nominal threshold of 5 are indicative of detector non-linearity or calibration degradation. These pixels are considered unreliable for accurate spectral analysis.}
\label{fig:phd_125_130_pj6}
\end{figure}

\begin{figure}[h]
\centering
\includegraphics[width=0.48\textwidth]{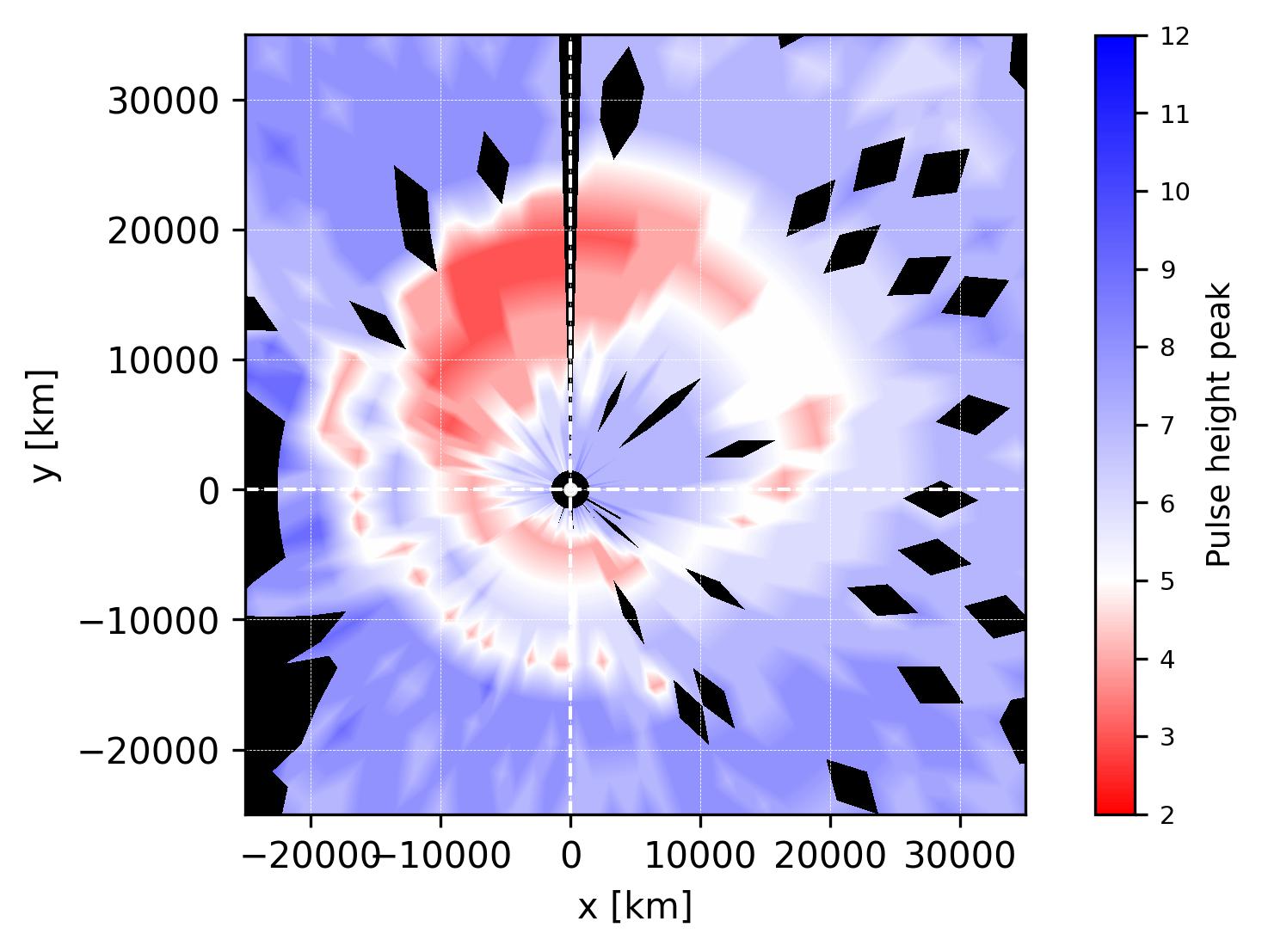}
\caption{Same as Fig.~\ref{fig:phd_125_130_pj6}, but for PJ10. The PHD degradation in the 125--130~nm interval is also prominent here.}
\label{fig:phd_125_130_pj10}
\end{figure}

\begin{figure}[h]
\centering
\includegraphics[width=0.48\textwidth]{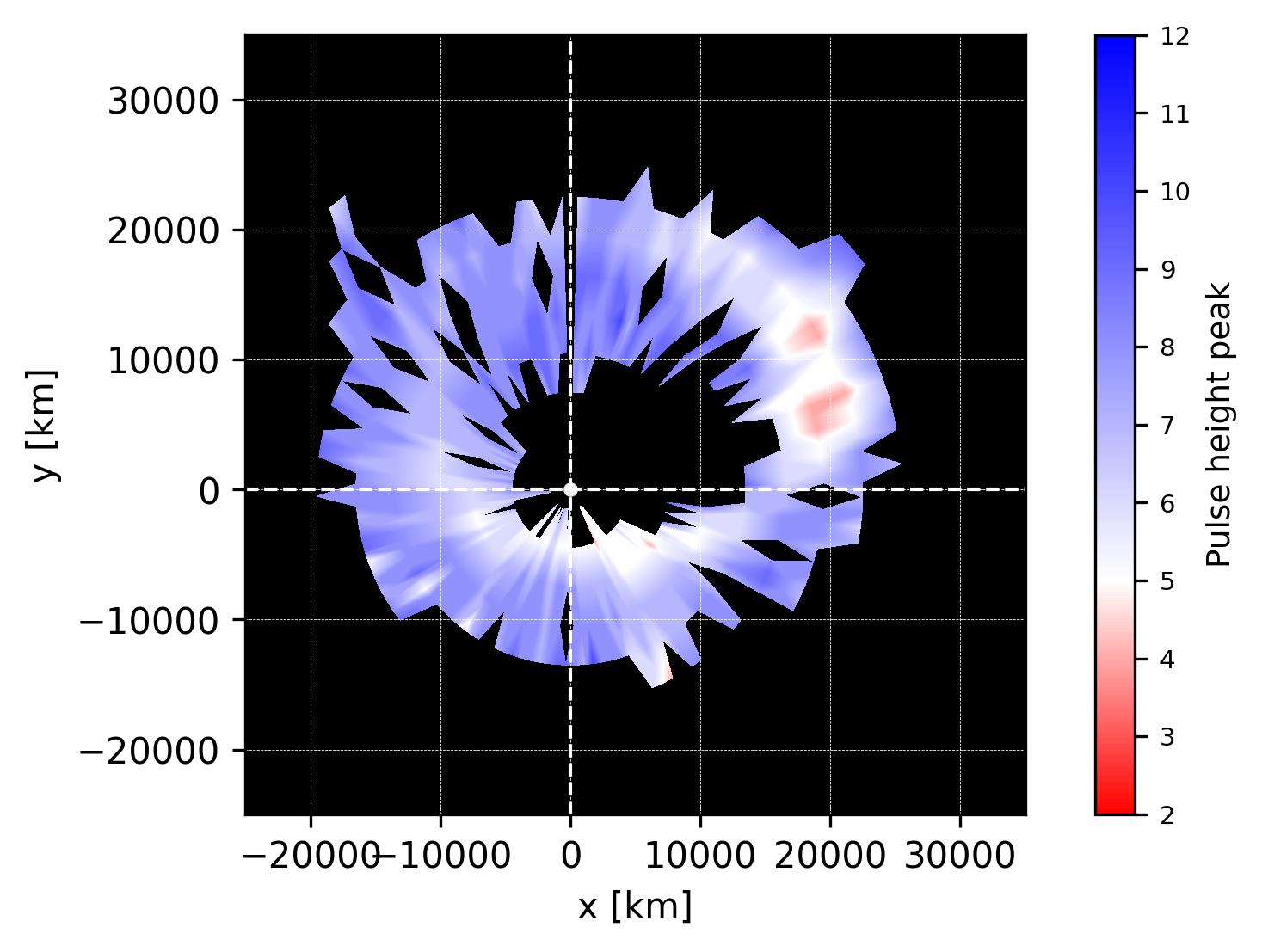}
\caption{PHD peak map for the 135--140~nm interval during the whole duration of PJ6. Most regions exhibit PHD peaks above the calibration threshold, indicating reliable spectral response.}
\label{fig:phd_135_140_pj6}
\end{figure}

\begin{figure}[h]
\centering
\includegraphics[width=0.48\textwidth]{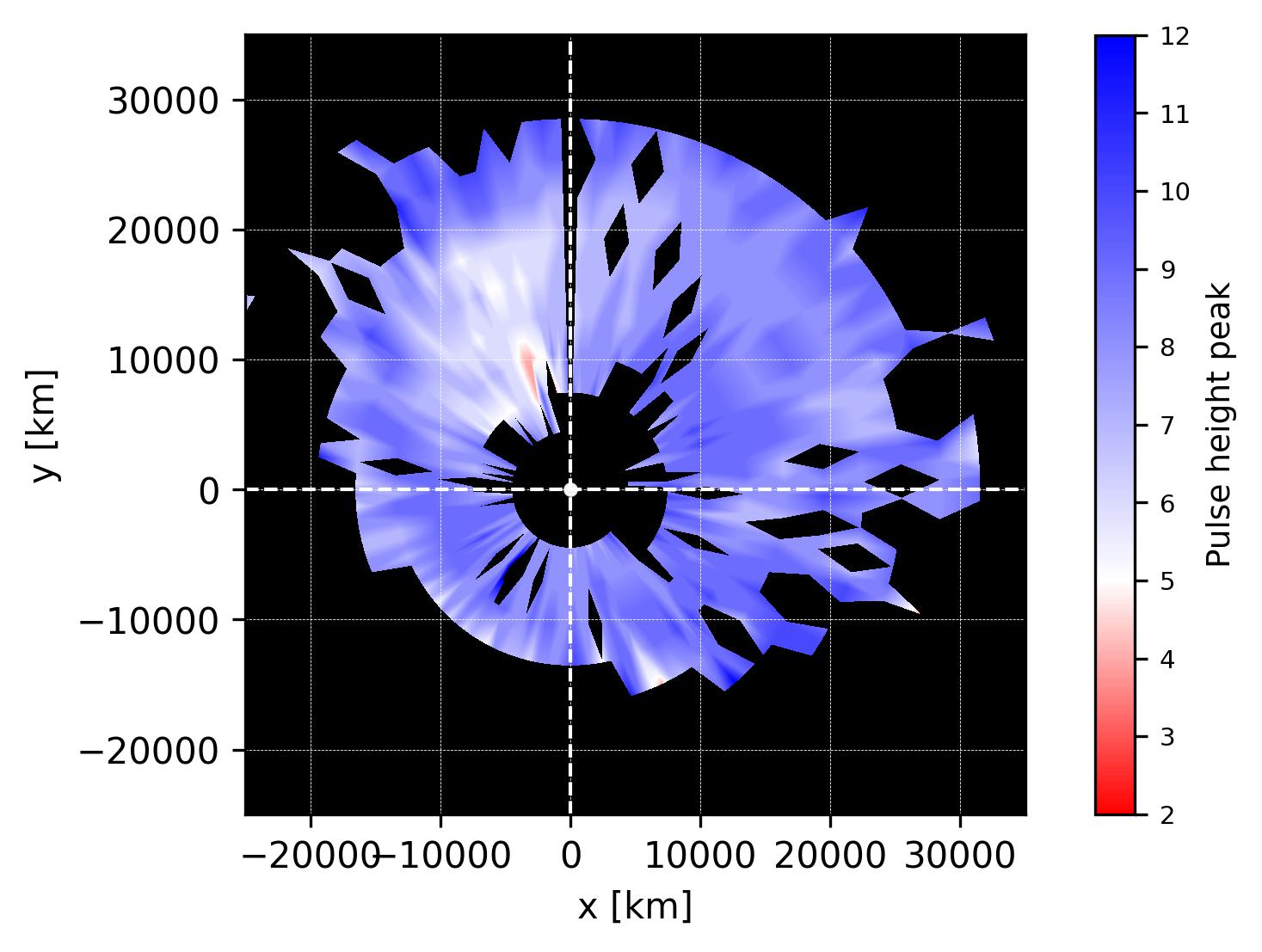}
\caption{Same as Fig.~\ref{fig:phd_135_140_pj6}, but for PJ10. The 135--140~nm interval shows consistent detector performance across the auroral region.}
\label{fig:phd_135_140_pj10}
\end{figure}

To mitigate these effects, several alternative intervals were tested and compared based on their correlation with the original CR and their intensity-to-noise ratio across multiple PJs (PJ1, PJ3, PJ6). Among these, the 135--140~nm band was found to offer the best compromise: 

\begin{itemize}
\item It lies within the CH$_4$ absorption range.
\item It displays a intensity-to-noise ratio close to or higher than the original 125-130~nm interval in all tested regions (see Table~\ref{tab:snr_compare}).
\item A comparative analysis of the PHD peak maps for PJ6 and PJ10 (Figs.~\ref{fig:phd_125_130_pj6}, \ref{fig:phd_125_130_pj10}, \ref{fig:phd_135_140_pj6}, \ref{fig:phd_135_140_pj10}) shows that the 125-130~nm interval frequently exhibits PHD peaks below the nominal calibration threshold of 5, whereas the 135-140~nm interval remains within expected ranges. This confirms that the latter is significantly less affected by instrumental distortions.
\end{itemize}

These conclusions were confirmed by comparing CR maps using different absorption intervals and by checking the PHD in regions where the correlation with the classical ratio broke down. In all cases, the 135--140~nm interval provided reliable and robust spectral behavior.

We thus adopted the following updated CH$_4$ CR definition for this study
\begin{equation}
\label{eq:newcr}
\mathrm{CR}_{\text{CH}_4} = \frac{I(155\text{--}162~\text{nm})}{I(135\text{--}140~\text{nm})}.
\end{equation}

This redefinition ensures that the CR remains sensitive to CH$_4$ absorption while minimizing instrumental biases due to non-linearity and PHD degradation.

\begin{table}[h]
\scriptfootnotesize

\centering
\caption{Intensity-to-noise ratio (INR) comparison for different candidate absorption intervals in five representative auroral regions. The 135--140~nm interval consistently shows the highest INR.}
\label{tab:snr_compare}
\begin{tabular}{@{}lccccc@{}}
\hline\hline
Interval (nm) & Region 1 & Region 2 & Region 3 & Region 4 & Region 5 \\
\hline
125--130   & 4.62 & 3.86 & 9.34 & 4.23 & 10.83 \\
130--140   & 4.24 & 1.75 & 5.02 & 2.44 & 5.70 \\
135--140   & \textbf{4.90} & \textbf{2.21} & \textbf{5.92} & \textbf{2.97} & \textbf{6.51} \\
130--135   & 3.19 & 1.13 & 2.60 & 1.49 & 4.21 \\
132--137   & 4.33 & 1.85 & 4.61 & 2.23 & 5.95 \\
\hline
\end{tabular}
\end{table}

\section{Spectral Flattening Detection and Correction Method}
\label{app:erosion}

The spectra gathered by Juno-UVS during PJ6 in the region of a dawn storm in the southern aurora are flattened and the CR maps that we construct from the spectral cube of this PJ can not be used as a diagnostic for hydrocarbon absorption. 
To work around the high-flux non-linearity effects that erode the spectral shape in high-brightness auroral regions, we develop a method to identify and correct the CR for pixels affected by this high-flux issue. Our approach relies on defining two narrow diagnostic bands within the unabsorbed portion of the H$_2$ emission spectrum. These bands are selected based on both their spectral position and their sensitivity to instrumental degradation.

\begin{itemize}
    \item \textbf{Band A:} 157–158.5\,nm — centered on the first of the two prominent peaks in the unabsorbed range of the H$_2$ spectrum. This feature arises from strong Werner band emissions and is highly sensitive to spectral flattening in saturated observations.
    \item \textbf{Band B:} 163–165\,nm — located in a region of relatively smooth continuum, beyond the peak emission zone. This band serves as a stable normalization reference that is less susceptible to degradation of the PHD.
\end{itemize}

\begin{figure}[h]
\centering
\includegraphics[width=0.5\textwidth]{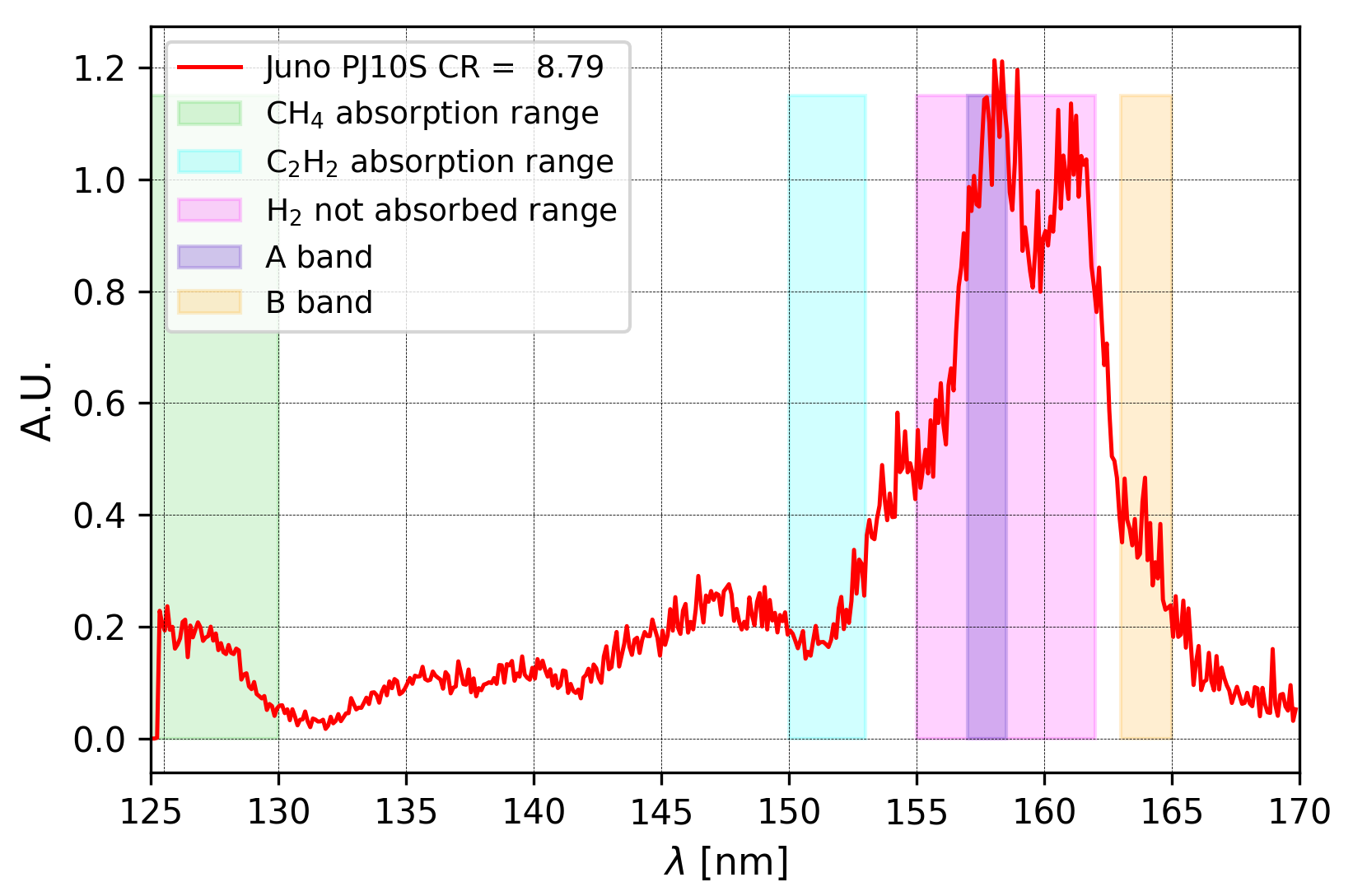}
\caption{Example of an H$_2$ auroral emission spectrum from PJ10, under non-saturated conditions. The diagnostic bands A (violet) and B (orange) are highlighted. Band A captures the first emission peak, while band B provides a stable reference point.}
\label{fig:diagnostic_bands}
\end{figure}

The motivation behind this selection is illustrated in Fig.~\ref{fig:diagnostic_bands}, which shows a representative H$_2$ auroral spectrum obtained from PJ10 under non-saturated conditions. Band A captures the height of a sharp spectral peak, making it a sensitive probe of any flattening due to gain compression. Band B, on the other hand, provides a reference baseline that is minimally affected by non-linearities and changes in absorption.

\begin{figure}[h]
\centering
\includegraphics[width=0.48\textwidth]{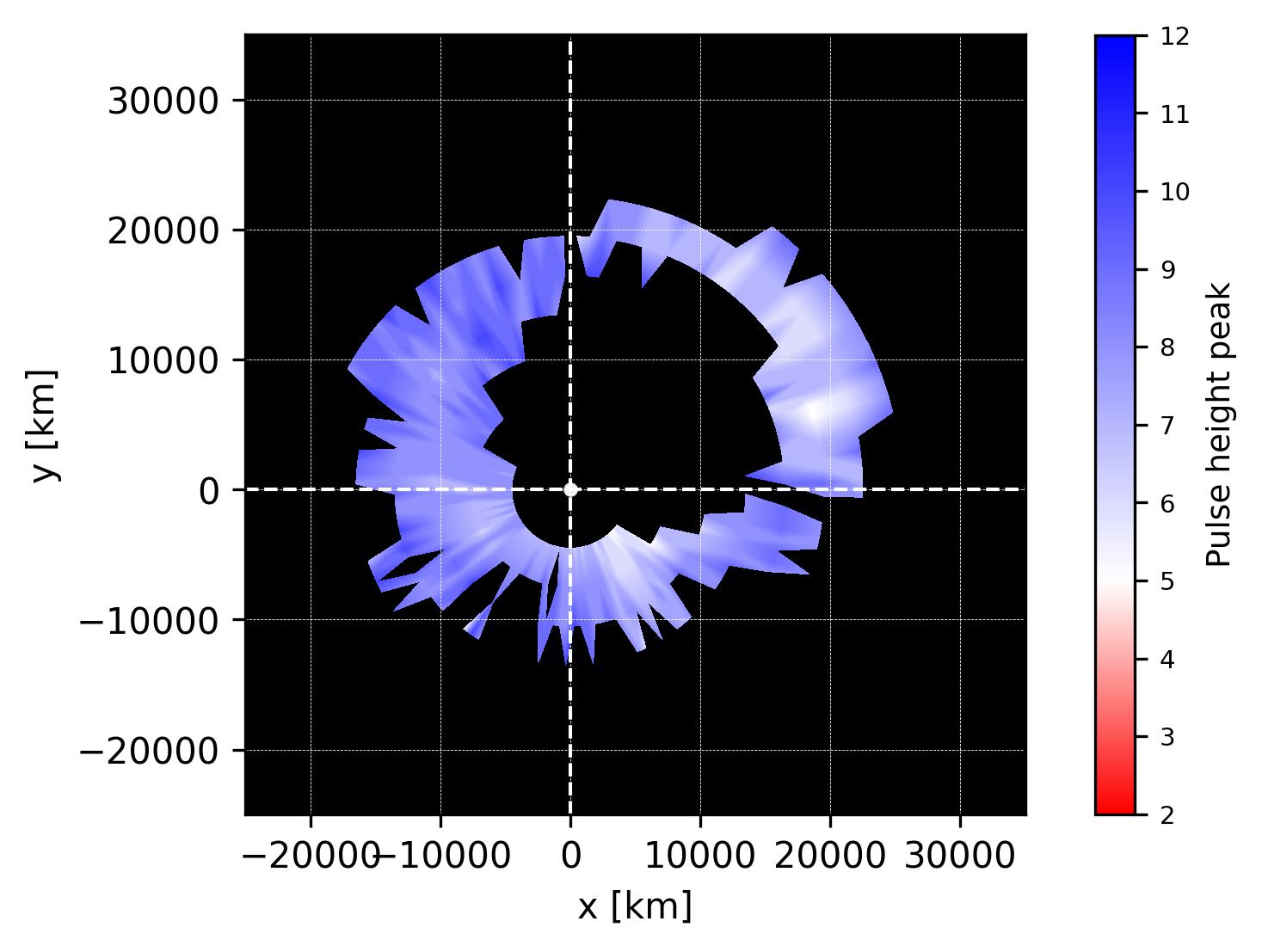}
\caption{PHD peak values in band B (163--165\,nm) during the whole duration of PJ6. The PHD remains within the nominal range across the entire auroral region, indicating that this spectral band is not significantly affected by detector non-linearity during PJ6.}
\label{fig:phd_map_bandb_pj6}
\end{figure}

\begin{figure}[h]
\centering
\includegraphics[width=0.48\textwidth]{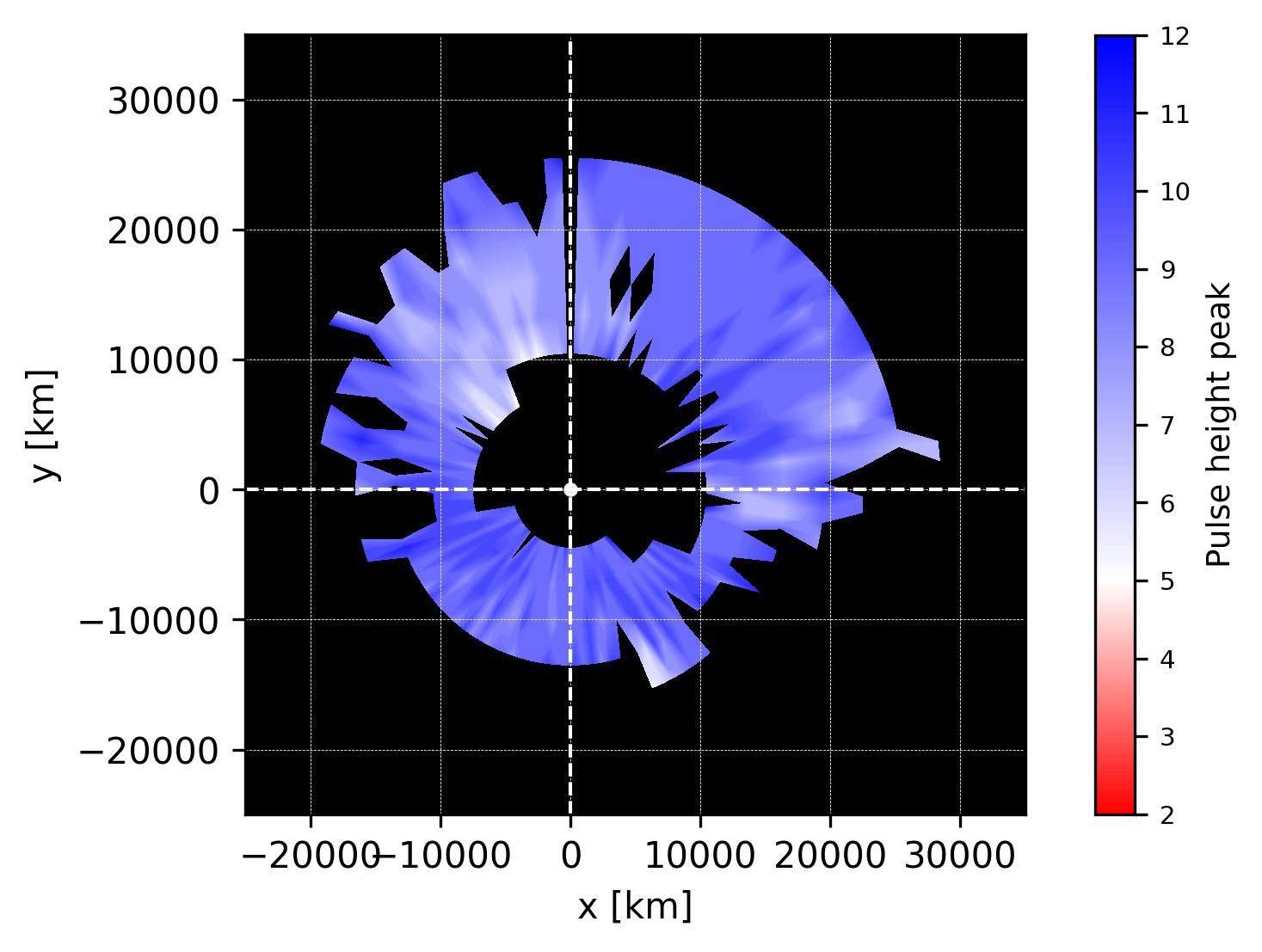}
\caption{Same as Fig.~\ref{fig:phd_map_bandb_pj6}, but for PJ10. Again, band B exhibits consistently calibrated PHD values.}
\label{fig:phd_map_bandb_pj10}
\end{figure}

Crucially, we also verified that the PHD in Band B remains systematically well-behaved, even in high-brightness regions where the unabsorbed 155–162\,nm band is significantly degraded. This is demonstrated in Figs.~\ref{fig:phd_map_bandb_pj6} and~\ref{fig:phd_map_bandb_pj10}, which show the PHD peak maps for band B during PJ6 and PJ10, respectively. In both cases, the PHD peak values remain above the nominal calibration threshold (typically set to 5), confirming that band B is less sensitive to detector high flux non-linearity than the broader unabsorbed interval.

\begin{figure}[h]
\centering
\includegraphics[width=0.48\textwidth]{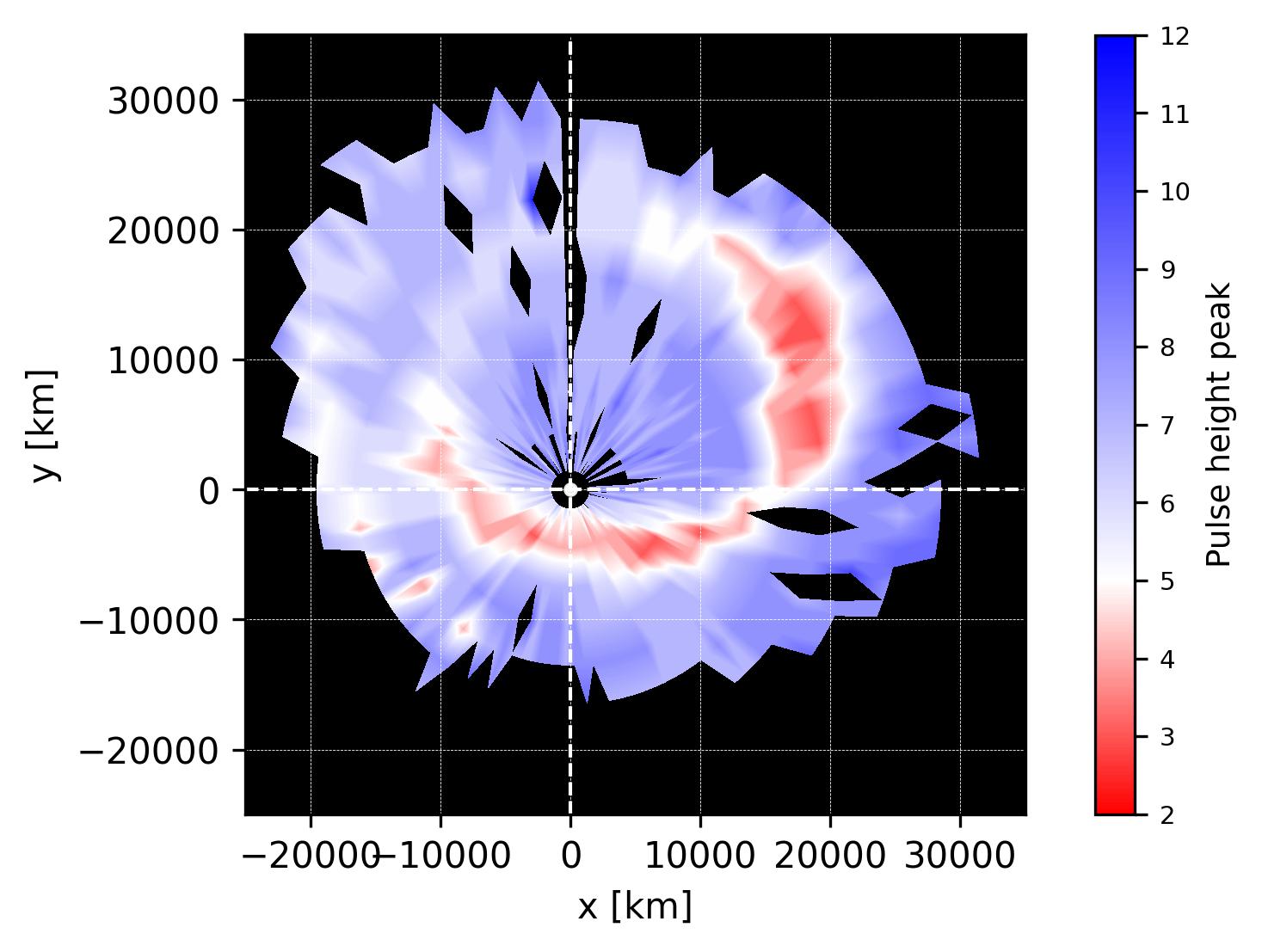}
\caption{PHD peak values in the 155--162\,nm band during the whole duration of PJ6. Regions with values below 5 (pink patches) are indicative of high flux and probable detector non-linearity, particularly within the brightest parts of the aurora.}
\label{fig:phd_map_155162_pj6}
\end{figure}

\begin{figure}[h]
\centering
\includegraphics[width=0.48\textwidth]{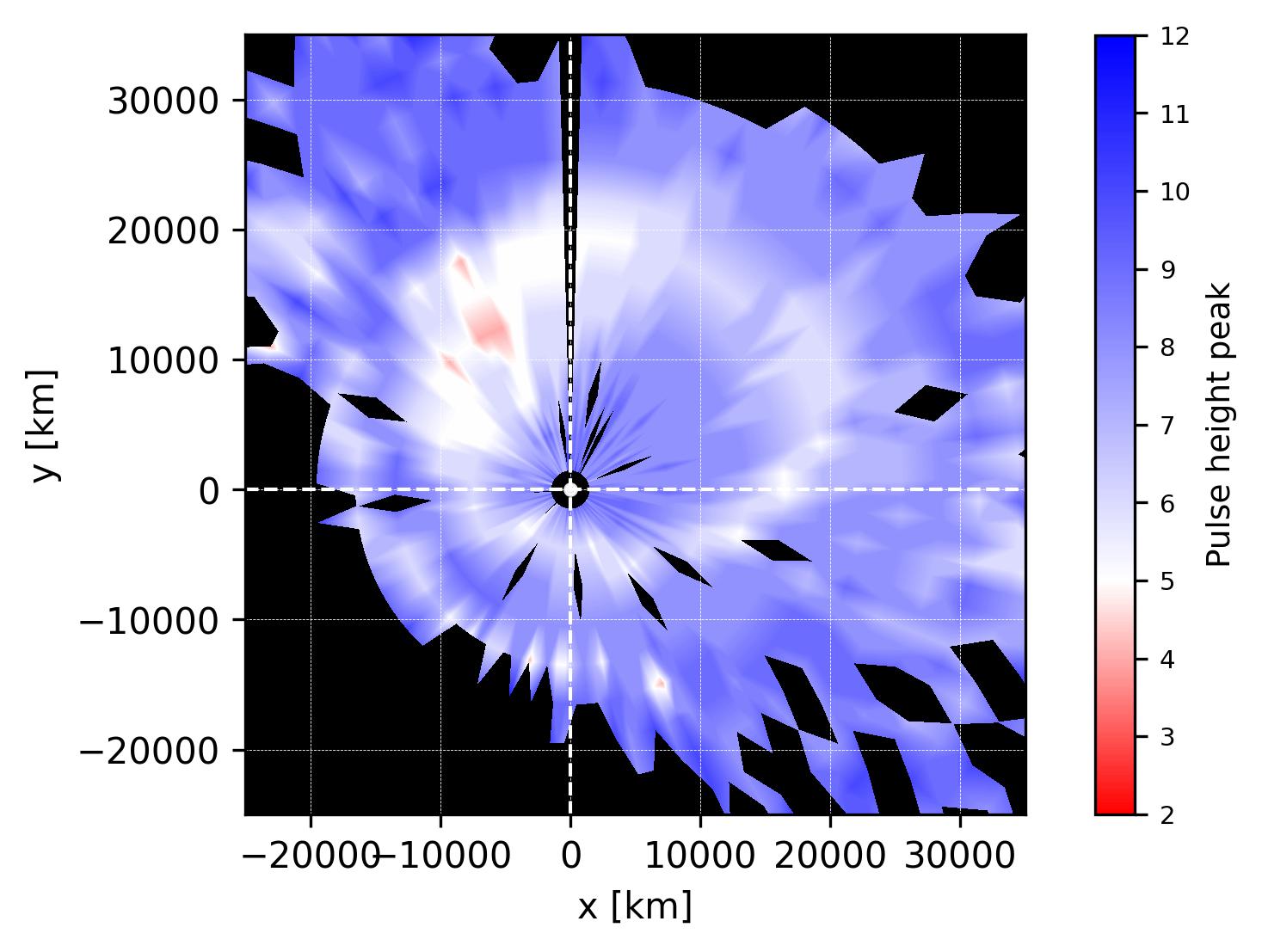}
\caption{Same as Fig.~\ref{fig:phd_map_155162_pj6}, but for PJ10. Spectral erosion is again apparent in the brightest auroral regions.}
\label{fig:phd_map_155162_pj10}
\end{figure}

For comparison, the corresponding PHD peak maps for the 155–162\,nm interval are shown in Figs.~\ref{fig:phd_map_155162_pj6} and~\ref{fig:phd_map_155162_pj10}. These maps reveal extended regions with degraded PHDs (peak values below 5), especially within the brightest parts of the aurora. This contrast justifies the use of band B as a more reliable normalization anchor in our correction methodology.

We use TransPlanet simulations to compute the theoretical A/B intensity ratio under different auroral precipitation conditions. Specifically, six simulations were performed for electron energies of 1, 45 and 85~keV at both poles. The A/B ratios obtained are summarized in Table~\ref{tab:bd_ratio}. These values vary little across simulations, confirming that the 163–165~nm (band A) and 170–172~nm (band B) regions are minimally affected by hydrocarbon absorption. We therefore adopt the average ratio of 2.4639 as a reference baseline.

\begin{table}[ht]
\centering
\caption{Theoretical A/B intensity ratios from TransPlanet simulations.}
\label{tab:bd_ratio}
\begin{tabular}{lccc|c}
\hline\hline
 & 1~keV & 45~keV & 85~keV & \multirow{2}{*}{Mean} \\
\cline{1-4}
A/B (North) & 2.4688 & 2.4602 & 2.4584 & \multirow{2}{*}{2.4639} \\
A/B (South) & 2.4697 & 2.4677 & 2.4585 & \\
\hline
\end{tabular}
\end{table}

Pixels exhibiting A/B ratios lower than 95\% of the theoretical mean value are classified as degraded, provided that the B-band brightness exceeds 0.4~kR to limit the influence of photon noise. This dual criterion enables a robust detection of spectral distortions linked to detector non-linearity in regions of high auroral brightness.

For pixels flagged as spectrally degraded, the correction procedure relies on the assumption that the B band (163--165~nm), located outside the main hydrocarbon absorption features, remains largely unaffected by detector non-linearity for the studied perijoves. The observed intensity in this reference band is therefore considered reliable. By applying the theoretical A/B ratio, derived from TransPlanet simulations, we can reconstruct the expected intensity in Band A (155--162~nm), which is sensitive to detector non-linearity.

This scaling allows us to estimate corrected CRs that more faithfully represent the true level of absorptions by CH$_4$ and C$_2$H$_2$. Specifically, the corrected CH$_4$ and C$_2$H$_2$ CRs are computed as

\begin{align}
    \mathrm{CR}_{\text{CH}_4} &= \frac{I(163\text{--}165~\text{nm})}{I(135\text{--}140~\text{nm})} \left( \frac{I(155\text{--}162~\text{nm})}{I(163\text{--}165~\text{nm})} \right)_{\text{th}} \\
    \mathrm{CR}_{\text{C}_2\text{H}_2} &= \frac{I(163\text{--}165~\text{nm})}{I(150\text{--}153~\text{nm})} \left( \frac{I(155\text{--}162~\text{nm})}{I(163\text{--}165~\text{nm})} \right)_{\text{th}}
\end{align}

This approach effectively compensates for the underestimation of flux in the unabsorbed band caused by detector non-linearity, while preserving the integrity of the spectral information in unaffected regions.

\end{document}